\documentclass[preprints,article,accept,moreauthors,10pt,a4paper]{mdpi}
\usepackage[utf8]{inputenc}
\usepackage{MnSymbol}
\firstpage{1} 
\pubvolume{xx}
\articlenumber{5}
\pubyear{2019}
\copyrightyear{2019}
\history{Received: 2 August 2019; Accepted: 1 October 2019; Published: date}

\Title{ Temperature Dependence of the Axion Mass in a Scenario Where the Restoration of Chiral Symmetry Drives the Restoration of the $U_A(1)$ Symmetry}

\Author{Davor Horvati\'c $^{1,2}$, Dalibor Kekez $^{3}$ and Dubravko Klabu\v{c}ar $^{1,*}$\orcidA{}}

\AuthorNames{Davor Horvati\'c, Dalibor Kekez and Dubravko Klabu\v{c}ar}

\address{%
$^{1}$ Physics Department, Faculty of Science-PMF, University of Zagreb, Bijeni\v{c}ka cesta 32, 
10000 Zagreb, Croatia \\
$^{2}$ davorh@phy.hr \\
$^{3}$ Ruđer Bo\v skovi\'c Institute, Bijeni\v{c}ka cesta 54, 10000 Zagreb, Croatia; kekez@irb.hr}

\corres{Correspondence: klabucar@phy.hr}

\abstract{The temperature ($T$) dependence of the axion mass is predicted
 for $T'$s up to $\sim 2.3 \times$ the chiral restoration temperature
 of QCD.  The axion is related to the $U_A(1)$ anomaly. The squared axion mass
 $m_{\rm a}(T)^2$ is, modulo the presently undetermined scale of
 spontaneous breaking of Peccei–Quinn symmetry $f_{\rm a}$ (squared),
 equal to QCD topological susceptibility $\chi(T)$ for all $T$. We obtain
 $\chi(T)$ by using quark condensates calculated in two effective
 Dyson–Schwinger models of nonperturbative QCD. They exhibit the correct
 chiral behavior, including the dynamical breaking of chiral symmetry and
 its restoration at high $T$. This is reflected in the $U_A(1)$ symmetry
 breaking and restoration through $\chi(T)$. In our previous studies,
 such $\chi(T)$ yields the $T$-dependence of the $U_A(1)$-anomaly-influenced 
 masses of $\eta'$ and $\eta$ mesons consistent with experiment. This in turn
 supports our prediction for the $T$-dependence of the axion mass.
 Another support is a rather good agreement with the pertinent lattice
 results. This agreement is not spoiled by our varying $u$ and $d$ quark
 mass parameters out of the isospin limit.  }


\begin{document}


\section{Introduction}

The axion, one of the oldest hypothetical particles beyond the Standard Model,
intensely sought for by many experimentalists already for 40 years now, still
escapes detection \cite{Tanabashi:2018oca}. It was introduced theoretically
\cite{Peccei:1977hh,Peccei:1977ur,Weinberg:1977ma,Wilczek:1977pj}
to solve the so-called Strong CP problem of QCD. The problem is that
no experimental evidence of CP-symmetry violation has been found in strong interactions, 
although the QCD Lagrangian $\,{\cal L}_{\rm QCD}(x)\,$ can
include the so-called $\theta$-term $\, {\cal L}_{\theta}(x)\, =\,\theta \, Q(x)\,$
where gluon field strengths $\, F^b_{\mu\nu}(x)\, $ form the CP-violating
 combination $Q(x)$ named the topological charge density:
\begin{equation}
Q(x)\, = \, \frac{g^2}{\, 32\,\pi^2\,}\, F^b_{\mu\nu}(x)\, \widetilde{F}^{b\mu\nu} \, ,
\qquad \mbox{where} \,\,\, 
\widetilde{F}^{b\mu\nu} \equiv \frac{1}{2} \,\epsilon^{\mu\nu\rho\sigma} \, F^b_{\rho\sigma}(x) .
\label{topolChargeDens}
\end{equation}
Whereas $Q(x)$ can be re-cast in the form of a total divergence
$\partial_\mu K^\mu$, discarding ${\cal L}_{\theta}$ is not justified
even if $\, F^b_{\mu\nu}(x)\, $ vanish sufficiently fast as $|x| \to \infty$. 
Specifically, $F\widetilde{F} = \partial_\mu K^\mu$ can anyway contribute to the
action integral, since in QCD there are topologically nontrivial field
configurations such as instantons. They are important for, {\it e.g.},
obtaining the anomalously large mass of the $\eta'$ meson. Also, precisely
the form (\ref{topolChargeDens}) from the $\theta$-term appears in the
 axial anomaly, breaking the $U_A(1)$ symmetry of QCD
 - see Equation~(\ref{divergenceOfA0mu}).

For these reasons, one needs 
$\, {\cal L}_{\theta}(x)\, =\,\theta \, Q(x)\,$
in the QCD Lagrangian, as reviewed briefly in Section 1 of Ref. \cite{Peccei:2006as}.
Moreover, the Strong CP problem cannot be removed by
requiring that the coefficient $\theta=0$, since QCD is an integral part
of the Standard Model, where  weak interactions break the CP symmetry.   
This CP violation comes from the complex Yukawa couplings, yielding the
 complex CKM matrix \cite{Cabibbo:1963yz,Kobayashi:1973fv} and
 the quark-mass matrix $M$ which is complex in general. 
To go to the mass--eigenstate basis, one diagonalizes the mass matrix,
and the corresponding chiral transformation changes $\theta$ by
$\arg\det M$. Hence, in the Standard Model the coefficient of the 
$Q \propto F\widetilde{F}$ term is in fact $\bar{\theta}=\theta+\arg\det M$
\cite{Peccei:2006as}. Therefore, to be precise, we change our notation to
$\bar{\theta}$-term, $\, {\cal L}_{\theta} \to \, {\cal L}_{\bar\theta}$.

Since CP is not a symmetry of the Standard Model, there is no {\it a priori}
 reason $\bar{\theta}$, which results from the contributions from {\it both}
 the strong and weak interactions, should vanish. 
 And yet, the experimental bound on it is extremely low,
 $|\bar{\theta}| < 10^{-10}$ \cite{Baker:2006ts},
 and in fact consistent with zero.  Therefore, the mystery of the vanishing
 strong CP violation is: {\it  why is  $\bar\theta$  so small? }

The most satisfactory answer till this very day has been provided by axions,
even though the original variant has been ruled out \cite{Tanabashi:2018oca}.
In the meantime, they turned out to be very important also for cosmology, as
promising candidates for dark matter---see from relatively recent
 references such as \cite{Wantz:2009it,Berkowitz:2015aua} to 
 the earliest papers \cite{Preskill:1982cy,Abbott:1982af,Dine:1982ah}.
(For an example of a broader review of axion physics, see \cite{Kim:2017tdk}.)
It is thus no wonder that ever since the original proposal of the axion
mechanism \cite{Peccei:1977hh,Peccei:1977ur,Weinberg:1977ma,Wilczek:1977pj}
in 1977--1978, many theorists kept developing various ideas on this
theoretically much needed object, trying to pinpoint the properties
of this elusive particle and increase chances of finding it.

However, to no avail.
There have even been some speculations that the axion is hidden in plain sight, by
being experimentally found, paradoxically, already years before it was conjectured
theoretically: namely, that the axion should in fact be identified with the
well-known $\eta'$ meson with a minuscule admixture of a pseudoscalar composite
of neutrinos \cite{Dvali:2016eay}. Nevertheless, while an intimate relation
 between the axion and $\eta'$ doubtlessly exists, reformulations of the
 axion theory, let alone so drastic ones, are in fact not needed to exploit
 this axion-$\eta'$ relationship: thanks to the fact that both of their masses
 stem from the axial anomaly and are determined by the topological susceptibility 
 of QCD, in the present paper we show how our previous study \cite{Horvatic:2018ztu}
 of the temperature ($T$) dependence of the $\eta'$ and $\eta$ mesons give us a
 spin-off in the form of the $T$-dependence of the axion mass, $m_{\rm a}(T)$.
 It is given essentially by the QCD topological susceptibility $\,\chi(T)\,$, which
 is rather sensitive to changes of the lightest quark masses $\, m_q $: Equation
 (\ref{chiLIGHTq}) vanishes linearly when $\, m_q\to 0 \,$ even for just one
 flavor $q$.
 We thus examine the effect of their values on $\chi(T)$ also out of the
 isosymmetric limit, and find that such a variation can be accommodated well. 
 The agreement with lattice results on $\chi(T)$ is reasonably good.

\section{Connection with the Complex of the $\eta'$ and $\eta$ Mesons}
\label{etaPeta}

\subsection{Some Generalities on the Influence of the Anomaly on $\eta'$ and $\eta$ }
\unskip

In this paper, we neglect contributions of quark flavors heavier than $q=s$
 and take $N_f=3$ as the number of active flavors.

At vanishing and small temperatures, $T\approx 0$, the physical $\eta'$ meson
is predominantly{\footnote{The mass eigenstate $\,\eta'\,$ is approximated only
roughly by the pure $SU(3)$ singlet state  $\eta_0$, due to the relatively large
explicit breaking of the flavor $SU(3)$ symmetry by much heavier $s$-quark:
$\, 2\, m_s/(m_u\, +\, m_d)\, = \, 27.3 \pm 0.7\,$ \cite{Tanabashi:2018oca}.}}
$\eta_0$, the singlet state of the flavor $SU(3)$ group, just like its
physical partner, the lighter isospin-zero mass eigenstate $\eta$ is
predominantly the octet state $\eta_8$. Unlike the $SU(3)$ octet states
$\pi, K$ and $\eta_8$, the singlet $\eta_0$ is precluded from being a
light (almost-)Goldstone boson of the dynamical breaking of the 
(only approximate) chiral symmetry of QCD (abbreviated as DChSB).
Namely $\eta_0$ receives a relatively large anomalous mass contribution
from the non-Abelian axial ABJ{\footnote{ABJ anomaly stands for names of
Adler, Bell, and Jackiw, as a reminder of their pioneering work on anomalies
 \cite{Adler:1969gk,Bell:1969ts} exactly half a century ago this year.}}
 anomaly, or gluon anomaly for short. An even better name for it is the
$U_A(1)$ anomaly, since it breaks explicitly the $U_A(1)$ symmetry of QCD
on the quantum level.

The breaking of $U_A(1)$ by the anomaly makes the flavor singlet ($a=0$)
 axial current of quarks,
$A_0^\mu(x) \, = \, \sum_{q=u,d,s} \, {\bar q}(x)\, \gamma^\mu \, \gamma_5\, q(x)$,
not conserved even in the chiral limit:
\begin{equation}
 \partial_\mu A^\mu_0(x) = {\rm i} \sum_{q=u,d,s} \, 2 \, m_q\,
{\bar q}(x)\, \gamma_5\, q(x) \, + \, 2 \, N_f \, Q(x)~, 
\qquad (N_f = 3)~,
\label{divergenceOfA0mu}
\end{equation}
unlike the corresponding octet currents $A_a^\mu(x)$, $a=1,2, ..., N_f^2-1$.
In the chiral limit, the 
current masses of non-heavy quarks all vanish, $m_q \to 0$ $(q=u,d,s)$,
but the divergence of the singlet current $A^\mu_0$ is not vanishing due
 to the $U_A(1)$ anomaly contributing no other but the topological charge
 density operator $Q(x)$ (\ref{topolChargeDens})---that is, precisely
 the quantity responsible for the strong CP problem.

 The quantity related to the $U_A(1)$-anomalous mass
in the $\eta'$-$\eta$ complex is the QCD topological susceptibility $\chi$,
\begin{equation}
\chi \,  =  \,
\int d^4x \; \langle 0|\, {\cal T} \, Q(x) \, Q(0) \, |0 \rangle \; ,
\label{chi}
\end{equation}
where $\cal T$ denotes the time-ordered product.

\begin{figure}[H]
\centering
\includegraphics[width=12 cm]{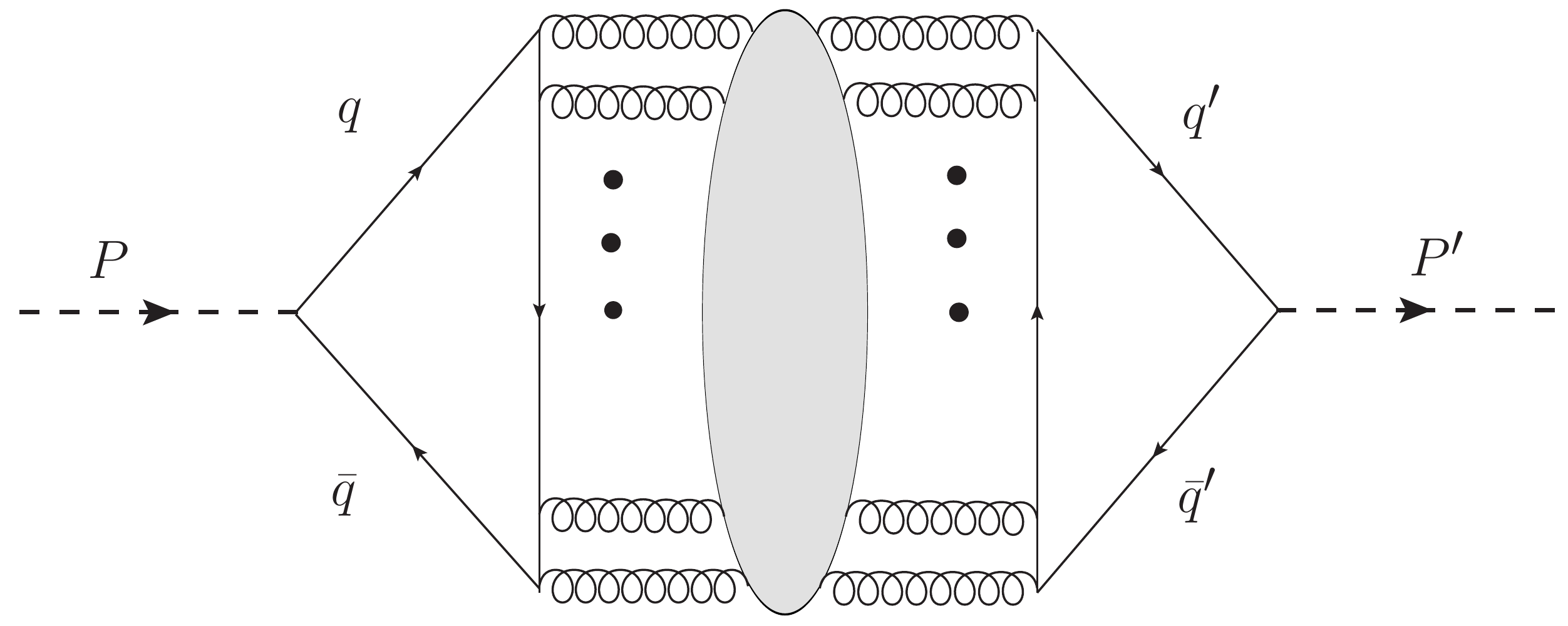}
\caption{$U_A(1)$ anomaly-induced, hidden-flavor-mixing transitions from
pseudoscalar quark–antiquark states $\, P=q\bar q\,$ to $\, P'=q'\bar q'\,$
include both possibilities $q=q'$ and $q\neq q'$. Springs symbolize gluons.
All lines and vertices are dressed in accord with the nonperturbative QCD.
Nonperturbative configurations are essential for nonvanishing anomalous mass
\cite{Feldmann:1999uf} contribution to $\eta_0 \sim \eta'$, since $Q(x)$ is
a total divergence. The gray blob symbolizes the infinity of all intermediate
gluon states enabling such transitions, so that the three bold dots represent
any  even \cite{Kekez:2000aw} number of additional gluons. Just one of
infinitely many, but certainly the simplest realization thereof, is when such
a transition is mediated by just two gluons (and no additional intermediate
states), whereby the above figure reduces to the so-called ``diamond graph''.
 }
\label{Blob}
\end{figure}

Figure \ref{Blob} shows how anomaly contributes to the mass matrix (in the
basis of quark–antiquark ($q\bar q$) pseudoscalar bound states $P$) by
depicting how hidden-flavor  $q\bar q$   pseudoscalars mix, transiting
 through anomaly-dominated gluonic intermediate states.

\subsection{On Some Possibilities of Modeling the $U_A(1)$ Anomaly Influence }
\unskip

Light pseudoscalar mesons can be studied by various methods. We have preferred using
\cite{Kekez:1996np,Kekez:1998xr,Klabucar:1997zi,Kekez:1998rw,Bistrovic:1999dy,Kekez:2000aw,Kekez:2001ph,Kekez:2003ri,Kekez:2005ie,Horvatic:2007mi,Horvatic:2007qs,Horvatic:2007wu,Horvatic:2018ztu}
 the relativistic bound-state approach to modeling nonperturbative QCD through
 Dyson–Schwinger equations (DSE),
 where, if approximations are consistently formulated, model DSE
 calculations also reproduce the correct chiral behavior of QCD. This is
 of paramount importance for descriptions of the light pseudoscalar mesons,
 which are quark–antiquark bound states but simultaneously also the
 (almost-)Goldstone bosons of DChSB of QCD. (For general reviews of the DSE approach,
 see, {\it e.g.}, Refs. \cite{Alkofer:2000wg,Roberts:2000aa,Holl:2006ni,Fischer:2006ub}.
 About our model choice at $T=0$ and $T>0$, in the further text see especially
 the Appendix.)

 Figure \ref{Blob} illustrates how hard computing would be ``in full glory'' the
 $U_A(1)$-anomalous mass and related quantities, such as the
 presently all-important topological susceptibility (\ref{chi})
 in the DSE approach with realistically modeled QCD interactions -- especially
 if the calculation should be
 performed in a consistent approximation with the calculation of the light
 pseudoscalar bound states, to preserve their correct chiral behavior.
 (For this reason, they have most often been studied in the rainbow-ladder
 approximation of DSE, which is inadequate for the anomalous contributions
 \cite{Alkofer:2000wg,Roberts:2000aa,Holl:2006ni,Fischer:2006ub},
 as also Figure \ref{Blob} shows.)

 However, our DSE studies of pseudoscalar mesons have been able to
 address not only pions and kaons, but also $\eta'$ and $\eta$ mesons, for
 which it is essential to include the anomalous $U_A(1)$ symmetry breaking
 at least at the level of the masses. This was done as described in Refs.
\cite{Klabucar:1997zi,Kekez:2000aw,Kekez:2005ie,Horvatic:2007mi,Benic:2014mha},
 namely exploiting the fact that the $U_A(1)$ anomaly is suppressed in the limit
 of large number of QCD colors $N_c$ \cite{Witten:1979vv,Veneziano:1979ec}.
 This allows treating the anomaly contribution formally as a perturbation
 with respect to the non-anomalous contributions to the $\eta$ and $\eta'$
 masses \cite{Klabucar:1997zi,Kekez:2000aw,Kekez:2005ie}. This way we avoid
 the need to compute the anomalous mass contribution together, and consistently,
 with the non-anomalous, chiral-limit-vanishing parts of the masses.
 The latter must be evaluated by some appropriate,
 chirally correct method, and our preferred tool---the relativistic bound-state
 DSE approach \cite{Alkofer:2000wg,Roberts:2000aa,Holl:2006ni,Fischer:2006ub}---is just one such possibility. The point is that they comprise the non-anomalous
 part of the $\eta'$-$\eta$ mass matrix, to which one can add, as a first-order
 perturbation, the $U_A(1)$-anomalous mass contribution $M_{U_A(1)}$---and it
 does not have to be modeled, but taken from lattice QCD \cite{Kekez:2005ie}.
 Specifically, at $T=0$, $M_{U_A(1)}$ can be obtained from $\chi_{\rm YM}$, the
 topological susceptibility of the (pure-gauge) Yang-Mills theory, for which
 reliable lattice results have already existed for a long time\footnote{This is
 in contrast with $\chi = \chi_{\rm QCD}$, the full-QCD topological susceptibility
 (\ref{chi}), which is much harder to find on the lattice because of the light
 quark flavors. $\chi = \chi_{\rm QCD}$ approaches $\chi_{\rm YM}$ only if one
 takes the quenched limit of infinitely massive quarks,
 $\chi_{\rm YM} = \chi_{\rm quench}$, since quarks then disappear from the
 loops of Equation (\ref{chi}).} \cite{Alles:1996nm,Boyd:1996bx,Gattringer:2002mr}.

This can be seen from the remarkable Witten–Veneziano relation (WVR)
 \cite{Witten:1979vv,Veneziano:1979ec}
 which in a very good approximation relates the full-QCD quantities ($\eta'$,
 $\eta$ and $K$-meson masses $M_{\eta'}, M_{\eta}$ and $M_K$ respectively, and
 the pion decay constant $f_\pi$), to the pure-gauge quantity $\chi_{\rm YM}$:
\begin{equation}
M_{\eta'}^2 \, + \, M_\eta^2 \, - \, 2 \, M_K^2 \, = \, 
   2 N_f \,\, \frac{\chi_{\rm YM}}{f_\pi^2} \, \equiv \, M_{U_A(1)}^2  \, .
\label{WittenVenez}
\end{equation}
The right-hand-side must be the total $U_A(1)$-anomalous mass contribution
in the $\eta'$-$\eta$ complex, since in the combination on the left-hand-side
everything else cancels at least to the second order, ${\cal O}({m}_q^2)$,
in the current quark masses of the three light flavors $q=u,d,s$.
This is because the non-anomalous, chiral-limit-vanishing parts $M_{q\bar q'}$
of the masses of pseudoscalar mesons\footnote{The combinations $P\sim q{\bar q}'$
 need not always pertain to physical mesons. The pseudoscalar hidden-flavor
 states $u\bar{u}$, $d\bar{d}$, $s\bar{s}$ are not physical as long as the
 $U_A(1)$ symmetry is not restored ({\it i.e.}, the anomaly effectively
 turned off, see around Equation (2.6) in Ref. \cite{DiVecchia:2017xpu} for example),
 but build the $SU(3)$ states $\eta_0$, $\eta_8$ and $\pi^0$.}
$P \sim q{\bar q}'$ composed of sufficiently light quarks, satisfy the
Gell–Mann–Oakes–Renner (GMOR) relation with their decay constants $f_{q\bar q'}$
 and the quark–antiquark ($q{\bar q}$) condensate signaling DChSB: 
\begin{equation}
M_{q\bar q'}^2 = \frac{-\langle {\bar q}\, q\rangle_0}{(f_{q\bar q'}^{\rm ch.lim})^2}
 \, ({m}_q + {m}_{q'}) \, + \, {\cal O}({m}_{q'}^2, {m}_q^2)\,
\qquad (q,q'=u,d,s) \, .
\label{GMOR}
\end{equation}
Here $f_{q\bar q'}^{\rm ch.lim} = f_{q\bar q'}({m}_q,{m}_{q'}\to 0)$, and 
 $\langle {\bar q}\, q\rangle_0$ denotes the massless-quark condensate, {\it i.e.},
 the $q{\bar q}$ chiral-limit condensate, or ``massless'' condensate for short.
(In the absence of electroweak interactions, the ``massless'' condensates have equal
values for all flavors: $\langle {\bar q}\, q\rangle_0 = \langle {\bar q'}q'\rangle_0$.)
It turns out that even $s$-flavor is sufficiently light
for Equation (\ref{GMOR}) to provide reasonable approximations.

 Using WVR and $\chi_{\rm YM}$ to get the anomalous part of the $\eta'$ and $\eta$
 masses is successful \cite{Kekez:2005ie} only for $T \sim 0$, or at any rate,
 $T$'s well below $T_c$, the pseudocritical temperature of the chiral transition.
 In the absence of a systematic re-derivation of WVR (\ref{WittenVenez}) at $T>0$,
 its straightforward extension (simply replacing all quantities by their $T$-dependent
 versions) is tempting, but was found \cite{Horvatic:2007qs} unreliable and with predictions
 in a drastic conflict with experiment when $T$ starts approaching $T_c$. This is because
 the full-QCD quantities $M_{\eta'}(T)$, $M_{\eta}(T)$, $M_K(T)$ and $f_\pi(T)$ have
 very different $T$-dependences from the remaining quantity $\chi_{\rm YM}(T)$, which
 is pure-gauge and thus much more resilient to increasing temperature: the critical
 temperature of the pure-gauge, Yang-Mills theory, $T_{YM}$, is more than 100 MeV
 higher than QCD's $T_c = (154\pm 9)$ MeV \cite{Dick:2015twa,Bazavov:2017dus}.
 The early lattice result $T_{YM}\approx 260$ MeV \cite{Alles:1996nm,Boyd:1996bx} is
 still accepted today \cite{Bazavov:2013txa}, and lattice groups finding a different
 $T_{YM}$ claim only it is even higher, for example $T_{YM}=(300\pm 3)$ MeV of
 Gattringer {\it et al.} \cite{Gattringer:2002mr}. (There are even some claims
 about experimentally established $T_{YM}=270$ MeV \cite{Stoecker:2015zea}.)

We thus proposed in 2011. \cite{Benic:2011fv} that the above mismatch
 of the $T$-dependences in WVR (\ref{WittenVenez}) can be removed if
one invokes another relation between $\chi_{\rm YM}$ and full-QCD
quantities, to eliminate $\chi_{\rm YM}$, {\it i.e.}, substitute
pertinent full-QCD quantities instead of $\chi_{\rm YM}$ at $T>0$.
This is the Leutwyler–Smilga (LS) relation, {Equation (11.16) of Ref.}
\cite{Leutwyler:1992yt}, which we used \cite{Benic:2011fv,Benic:2014mha,Horvatic:2018ztu}
in the inverted form (and in our notation):
\begin{equation}
\label{LS}
 \chi_{\rm YM} \, = \, 
\frac{\chi}{\,\, 1 + \, \chi \, (\,\frac{1}{ m_u} + \frac{1}{m_d} + \frac{1}{m_s} \,)
\, \frac{1}{\langle{\bar q}\, q\rangle_0} \,\, } \, \,
\,\, (\, \equiv \, {\widetilde \chi} \,\,) \,\, ,
\end{equation}
to express (at $T=0$) pure-gauge $\,\chi_{\rm YM}\,$ in terms of the full-QCD
topological susceptibility $\,\chi \equiv \chi_{\rm QCD}$, the current quark masses $m_q$,
and $\langle{\bar q}q\rangle_0$,
the condensate of massless, chiral-limit quarks. The combination which these full-QCD
quantities comprise, {\it i.e.}, the right-hand-side of the LS relation (\ref{LS}), we
denote (for all $T$) by the new symbol $\,\,{\widetilde \chi}\,\,$ for later convenience
- that is, for usage at high $T$, where the equality (\ref{LS}) with $\,\chi_{\rm YM}\,$
 does not hold.

 The remarkable LS relation (\ref{LS}) holds for
 all values of the current quark masses. In the limit of very heavy quarks, it
correctly yields $\chi \to \chi_{\rm quenched}=\chi_{\rm YM}$ for $m_q\to\infty$,
it but it also holds for the light $m_q$.  In the light-quark sector,
 the QCD topological susceptibility $\chi$ can be expressed as
 \cite{DiVecchia:1980yfw,Leutwyler:1992yt,Durr:2001ty}:
\begin{equation}
\chi = \frac{-\,\langle{\bar q}\, q\rangle_0}{\,\,\frac{1}{m_u}+\frac{1}{m_d}+\frac{1}{m_s}\,\,}
 \, + \, {\cal C}_m \, ,
\label{chi_small_m}
\end{equation}
where ${\cal C}_m$ represents corrections of higher orders in light-quark masses $m_q$.
Thus, it is small and often neglected, leaving just the leading term as the widely used
 \cite{Bernard:2012fw} expression for $\chi$ in the light-quark, $N_f=3$ sector.
However, setting ${\cal C}_m = 0$ in the light-quark $\chi$ (\ref{chi_small_m})
 returns us
 $\chi_{\rm YM}=\infty$ through Equation (\ref{LS})\cite{Benic:2011fv}. 
 Or conversely, setting $\chi_{\rm YM}=\infty$ in the LS  relation (\ref{LS}), gives
 the leading term of $\chi$ (\ref{chi_small_m}) (see also Ref. \cite{DiVecchia:2017xpu}).
 This can be a reasonable, useful limit
 considering that in reality $\chi_{\rm YM}/\chi \gtrsim 40$. Nevertheless, in our previous works
 on the $\eta'$-$\eta$ complex at $T>0$ \cite{Benic:2011fv,Horvatic:2018ztu}
 we had to fit ${\cal C}_m$ (and parameterize it with {\it Ans\"atze} at $T>0$),
 since we needed the realistic value of $\chi_{\rm YM}(T=0)={\widetilde \chi}(T=0)$
 from lattice to reproduce the well-known masses of $\eta$ and  $\eta'$ at $T=0$.
 (However, just for $\chi(T)$ this is not necessary.)

 Replacing \cite{Benic:2011fv} $\chi_{\rm YM}(T)$ by the full-QCD quantity
 $\widetilde\chi(T)$ obviously keeps WVR at $T=0$, but avoids the 'YM {\it vs.} QCD'
 $T$-dependence mismatch with $f_\pi(T)$ and the LHS of Equation (\ref{WittenVenez}),
so it is much more plausible to assume the straightforward extension of $T$-dependences.
 The $T$-dependences of $\chi(T)$ (\ref{chi_small_m}) and $\widetilde\chi(T)$, and thus
 also of the anomalous parts of the $\eta$ and $\eta'$ masses, are then obviously 
 dictated by $\langle{\bar q}q\rangle_0(T)$, the ``massless'' condensate.

\begin{figure}[htb]
\centerline{%
\includegraphics[width=14.2cm]{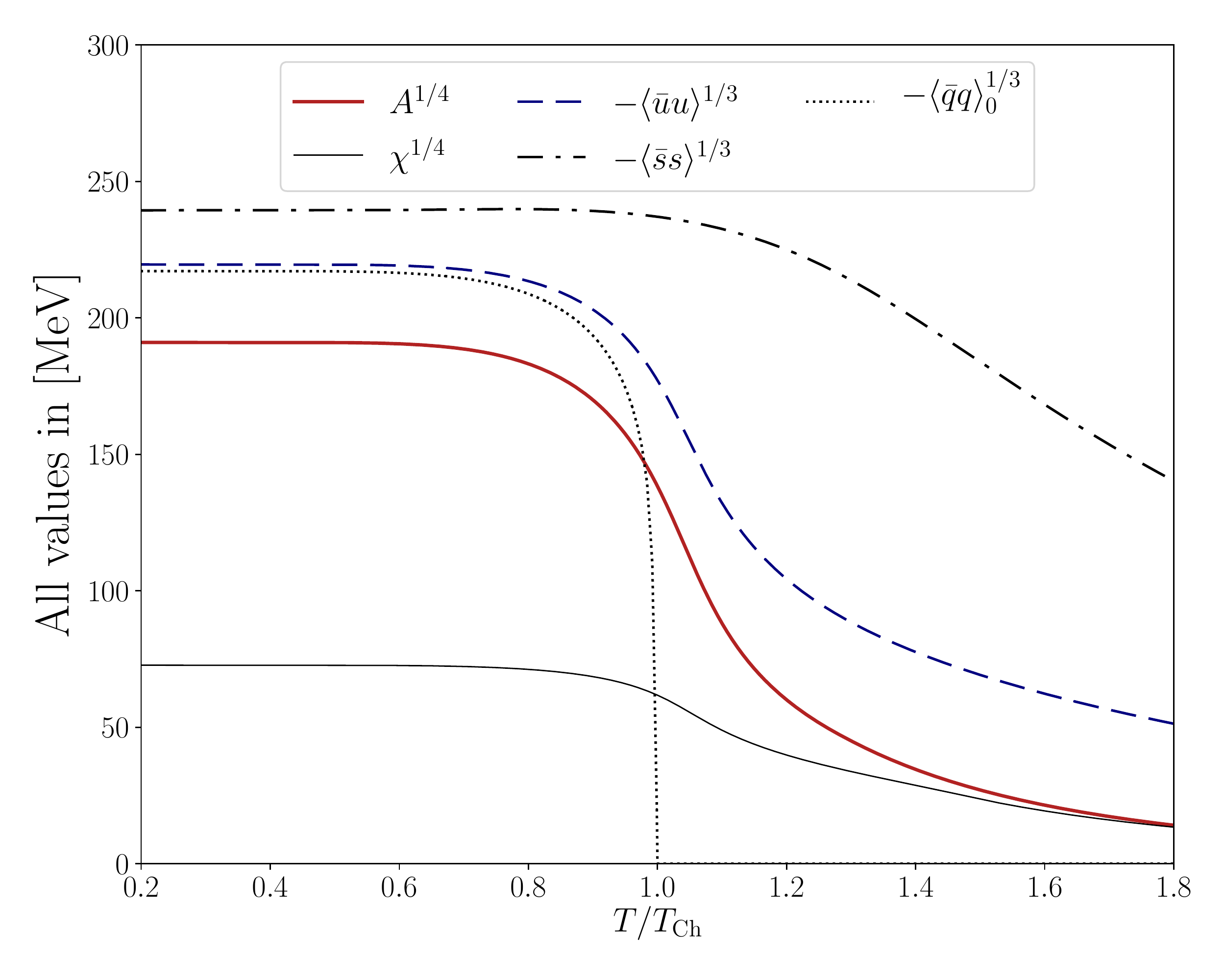}}
\caption{The relative-temperature $T/T_c$ dependences (where $T_{\rm Ch}\equiv T_c$)
of (the 3$^{\rm rd}$ root of the absolute value of) the $q\bar q$ condensates, and of
(the 4$^{\rm th}$ root of) the topological susceptibility $\chi(T)$ and the full
QCD topological charge parameter $A(T)$. Everything was calculated in {the isosymmetric
limit using} the separable rank-2 DSE model which we had already used in Refs.
 \cite{Horvatic:2007qs,Benic:2011fv,Horvatic:2018ztu}.
(See the Appendix for the model interaction form and parameters, {and the first line of
Table \ref{ourTable} for the numerical values of the condensates and $\chi$ at $T=0$.})
 Only the chiral-limit
condensate $\langle{\bar q}q\rangle_0(T)$ falls steeply to zero at $T=T_{\rm Ch}$,
indicative of the second-order phase transition. This sharp transition would through
(\ref{chi_small_m}) and (\ref{LS}) be transmitted to, respectively, the chiral-limit
$\chi(T)$ and ${\widetilde \chi}(T)$, and ultimately to the $\eta$ and $\eta'$
masses in Ref. \cite{Benic:2011fv}. The highest curve (dash-dotted)
and the second one from above (dashed) are (3$^{\rm rd}$ roots of the absolute values
of) the condensates $\langle{\bar s}s\rangle(T)$ and $\langle{\bar u}u\rangle(T)$,
respectively. Their smooth crossover behaviors carry over to $\chi(T)$ and $A(T)$
through Equations (\ref{chiLIGHTq}) and (\ref{defA}), respectively, leading to the
empirically acceptable predictions \cite{Horvatic:2018ztu} for the $T$-dependence
of the $\eta$ and $\eta'$ masses. It turns out that  $\chi(T)$ (\ref{chiLIGHTq}) also
 gives the smooth crossover behavior also to the $T$-dependence of the axion mass.}
\label{condens0uds}
\end{figure}

General renormalization group arguments suggest \cite{Pisarski:1983ms}
that QCD with three degenerate light-quark flavors has a first-order
phase transition in the chiral limit, whereas in QCD with (2+1) flavors
(where $s$-quark is kept significantly more massive) a second-order
chiral-limit transition{\footnote{This is a feature exhibited by DSE models, or at least by most of them,
through the characteristic drop of their chiral-limit, massless $q\bar q$ condensates
\cite{Holl:1998qs,Kiriyama:2001ah,Ikeda:2001vc,Blank:2010bz,Fischer:2009bh,Qin:2010nq,Gao:2015kea,Fischer:2018sdj}.}}
is also possible and even more likely \cite{Ejiri:2009ac,Ding:2018auz,Ding:2019fzc}.
What is important here, is that in any case the chiral-limit condensate
$\langle{\bar q}q\rangle_0(T)$ drops sharply to zero at $T=T_c$. (The
dotted curve in Figure \ref{condens0uds} is just a special example thereof,
namely $\langle{\bar q}q\rangle_0(T)$ calculated in Ref. \cite{Benic:2011fv}
using the same model as in \cite{Horvatic:2018ztu} and here.)
 This causes a similarly sharp drop of $\widetilde\chi(T)$ and $\chi(T)$.
 (We may be permitted to preview similar dotted curves in Figures \ref{FigIsosymm}
 and \ref{FigRank1all} in the next section and anticipate that in this case
 one would get a massless axion at $T=T_c$.) 
This was also the reason, besides the expected \cite{Csorgo:2009pa,Vertesi:2009wf}
 drop of the $\eta'$ mass, Ref. \cite{Benic:2011fv} also predicted so drastic drop
 of the $\eta$ mass at $T=T_c$, that it would become degenerate with the pion.
 However, no experimental indication whatsoever for a decreasing behavior of
 the $\eta$ mass, and much less for such a conspicuous sharp mass drop,
 has been noticed to this day,
 which seems to favor theoretical descriptions with a smooth crossover.
 Also, recent lattice QCD results (see
 \cite{Aoki:2012yj,Buchoff:2013nra,Dick:2015twa} and their references)
 show that the chiral symmetry restoration is a crossover around
 the pseudocritical transition temperature $T_c$.

%
%

To describe such a crossover behavior of the chiral transition, we
incorporated \cite{Benic:2014mha,Horvatic:2018ztu} into our approach
Shore's generalization \cite{Shore:2006mm} of the Witten–Veneziano relation.
To be precise, we studied it at $T=0$ already in 2008 \cite{Horvatic:2007mi}
and adapted it to our DSE bound-state context by applying some very plausible
simplifications \cite{Feldmann:1999uf}.
 The more recent reference \cite{Benic:2014mha} presented the analytic,
 closed-form solutions to Shore's equations for the pseudoscalar meson
 masses. These solutions showed that Shore's approach is then actually
 quite similar to the original WVR, leading to a similar $\eta$-$\eta'$
 mass matrix \cite{Benic:2014mha}.

Presently, the most important advantage is that Shore's generalization
leads to the crossover $T$-dependence.
As obtained in Section 3 for two specific interactions modeling the nonperturbative
QCD interaction, these condensates exhibit a smooth, crossover chiral symmetry
transition around $T_c$. Here Figure \ref{condens0uds} illustrates this generic
behavior by displaying the results obtained in Section 3 for a specific
DSE model: the higher current quark mass, the smoother the crossover behavior,
which then results in the crossover behavior also of other quantities,
like the presently all-important quantity, the QCD topological susceptibility
$\chi(T)$.

This comes about as follows:
the quantity which in Shore's mass relations \cite{Shore:2006mm} has the role of
$\chi_{\rm YM}$ in the Witten–Veneziano relation, is called the full-QCD topological
charge parameter $A$. (Shore basically took over this quantity from Di Vecchia and Veneziano
 \cite{DiVecchia:1980yfw}.) At $T=0$, it is approximately equal to $\chi_\text{YM}$ in the
 sense of $1/N_c$ expansion. Shore uses $A$ to express the QCD susceptibility $\chi$
 through a relation similar to the Leutwyler–Smilga relation (see Equation (2.11) and (2.12)
 in Ref. \cite{Shore:2006mm}), but using the condensates $\langle {\bar u}u \rangle$,
 $\langle{\bar d}d \rangle$, $\langle {\bar s}s \rangle$ of realistically massive
 $u,d,s$ quarks. The inverse relation, yielding $A$ (with the opposite sign convention),
 is the most illustrative for us:
\begin{equation}
\label{defA}
A \, = \, \frac{\chi}{\,\, 1 \, +\, {\chi} \, (\,\frac{1}{m_{u}\,\langle{\bar u}u\rangle}
 + \frac{1}{m_{d} \,\langle {\bar d}d \rangle}
+ \frac{1}{m_{s}\,\langle {\bar s}s \rangle } \,) \,} \,
 \qquad\quad  (\, A = \chi_\text{YM} + {\cal O}(\frac{1}{N_c}) \,\,\,\, \text{at} \,\,\, T=0 \,) .
\end{equation}
Obviously, it is analogous to the inverted LS relation (\ref{LS}) defining
$\,{\widetilde \chi}\,$, except that $A$ is expressed through ``massive'' condensates.
(If they are all replaced by $\langle {\bar q}q \rangle_0$, then $A\to {\widetilde \chi}$.)
They are in principle different for each flavor, but in the limit of usually
excellent isospin symmetry, $\langle{\bar u}u\rangle = \langle {\bar d}d \rangle$.

One can examine the limiting assumption $A = \infty$
in analogy with taking the limit $\chi_\text{YM}=\infty$ compared to $\chi =\chi_{QCD}$.
Then, for $A = \infty$ (be it in our Equation (\ref{defA}) or Shore's Equations (2.11) and
(2.12) for $\chi$), one recovers the leading term of the QCD topological
susceptibility expressed by the ``massive'' condensates.    
However, if one needs a finite $A$, as in $\eta$-$\eta'$ calculations \cite{Horvatic:2018ztu}
where one needs to reproduce $A \approx \chi_{YM}$, one also needs the appropriate
correction term ${\cal C'}_m$, just as ${\cal C}_m$ in Equation (\ref{chi_small_m}), so that:
\begin{equation}
\chi(T) \, = \,  \frac{ - \, 1}{ \,\,
             \frac{1}{\, m_{u} \, \langle {\bar u}u \rangle(T) } +
          \frac{1}{\, m_{d}\,\langle{\bar d}d  \rangle(T)} +
        \frac{1}{\, m_{s}\,\langle {\bar s}s  \rangle(T) }
            \,\, }   \, + \, {\cal C}_m'  \, .
\label{chiLIGHTq}
\end{equation}
Again, ${\cal C'}_m$ is a very small correction term of higher orders in
 the small current quark masses $m_q$ ($q=u,d,s$), and  we can neglect it
 in the present context, where we actually have a simpler task than finding
 $T$-dependence of the $\eta$ and $\eta'$ masses in Ref. \cite{Horvatic:2018ztu}.
 Since it turns out that for determining the $T$-dependence of the mass of
 the QCD axion we do not need to find $A$, we set ${\cal C'}_m = 0$ in this paper
 throughout. One needs just the topological susceptibility $\chi(T)$ for
 that, and just the leading term of (\ref{chiLIGHTq}) will suffice to
 yield the crossover behavior found on lattice ({\it e.g.}, in Refs.
 \cite{Petreczky:2016vrs,Borsanyi:2016ksw,Bonati:2015vqz}).

\section{ The Axion Mass from the Non-Abelian Axial Anomaly of QCD }


Peccei and Quinn (PQ) introduced \cite{Peccei:1977hh,Peccei:1977ur} a new global
 symmetry $U(1)_{\rm PQ}$ which is broken spontaneously at some very large, but
 otherwise still unknown scale $f_{\rm{a}} > 10^8$ GeV
 \cite{Tanabashi:2018oca,Guth:2018hsa}, which determines the absolute value of
 the axion mass $m_{\rm{a}}$. Nevertheless, this constant 
 cancels from ratios such as $m_{\rm{a}}(T)/m_{\rm{a}}(0)$, where $T$
 is temperature. {Thus}, useful insights and applications, such as those involving
 the nontrivial part of axion $T$-dependence, are possible in spite
 of $f_{\rm{a}}$ being presently unknown.

The factor in the axion mass which carries the nontrivial $T$-dependence, is
the QCD topological susceptibility $\chi(T)$. This quantity is also essential
for our description of the $\eta'$-$\eta$ complex at $T > 0$, since it relates
the $T$-dependence of the anomalous breaking of $U_A(1)$ symmetry.


\subsection{The Axion as the Almost-Goldstone Boson of the Peccei–Quinn Symmetry}
\unskip

The pseudoscalar axion field  $\mbox{\large a}(x)$ arises as the
(would-be massless) Goldstone boson of the spontaneous breaking of the
 PQ symmetry $U(1)_{\rm PQ}$ \cite{Weinberg:1977ma,Wilczek:1977pj}. 
The axion contributes to the total Lagrangian its kinetic term and its
interaction with fermions of the Standard model. Nevertheless,
what is important for the resolution of the strong CP problem, is that
the axion also couples to the topological charge density operator $Q(x)$
defined in Equation (\ref{topolChargeDens}) and generating the $U_A(1)$-anomalous
term in Equation (\ref{divergenceOfA0mu}). The $\bar\theta$-term
 in ${\cal L}_{QCD}$ thus changes into
\begin{equation}
{\cal L}_{\bar\theta} \to 
{\cal L}^{\bar\theta+}_{\rm  axion}  \,  
 = \, \left(\, \bar{\theta} \, + \frac{\mbox{\large a}}{f_{\rm a}} \right) \,
 \frac{g^2}{64\pi^2} \,
                 \epsilon^{\mu\nu\rho\sigma} F^b_{\mu\nu} F^b_{\rho\sigma} \, .
\label{theta+axion}
\end{equation}
 Because of this axion-gluon coupling, the $U(1)_{\rm PQ}$ symmetry
 is also broken {\it explicitly} by the $U_A(1)$ anomaly (gluon axial anomaly).
 This gives the axion a nonvanishing mass, $m_{\rm a} \neq 0$
 \cite{Weinberg:1977ma,Wilczek:1977pj}.

Gluons generate an effective axion potential, and its minimization leads to
the axion expectation value $\langle \mbox{\large a} \rangle$
which makes the modified coefficient of $Q(x)$ in Equation (\ref{theta+axion}) vanish:
$\,\bar{\theta}\, + 
{\langle\mbox{\large a}\rangle}/f_{\rm a}\,\equiv\, \bar{\theta}'\, =\, 0\,$.

Obviously, the experiments excluding the strong CP violation, such as 
\cite{Baker:2006ts}, have in fact been finding that consistent with zero
is $\bar{\theta}'$, the coefficient of $Q(x)$ in the QCD Lagrangian when 
 the axion exists.
The strong CP problem is thereby solved, irrespective of the initial value
 of $\bar\theta$. (Relaxation from any $\bar\theta$-value in the early Universe
 towards the minimum at
  $ \, \bar\theta \, = \, - \, \langle \mbox{\large a} \rangle/f_{\rm a}  $
is called misalignment production. The resulting axion oscillation energy is a good
candidate for cold dark matter
\cite{Wantz:2009it,Berkowitz:2015aua,Preskill:1982cy,Abbott:1982af,Dine:1982ah,Kim:2017tdk}.)


\subsection{Axion Mass from the Topological Susceptibility from Condensates of Massive Quarks}

Modulo the (squared) Peccei–Quinn scale $f_{\rm a}^2$, the axion mass squared is
at all temperatures $T$ given by the QCD topological susceptibility 
\cite{Berkowitz:2015aua,Petreczky:2016vrs,Borsanyi:2016ksw,Bonati:2015vqz,DiVecchia:2017xpu,Bonati:2018blm}
 very accurately \cite{diCortona:2015ldu,Gorghetto:2018ocs}
 (up to negligible corrections of the order $({pion \,\, mass})^2/f_{\rm a}^2$):
\begin{equation}
  m_{\rm a}^2(T) \,  =  \, \frac{1}{ f_{\rm a}^2} \, \chi(T) \, , 
\label{axionMass}
\end{equation}
as revealed by the quadratic term of the expansion of the effective axion potential.

\vskip 1mm

We explained in Section \ref{etaPeta} how the $U_A(1)$ symmetry-breaking
quantity $\chi(T)$ can be obtained through Equation (\ref{chiLIGHTq}) as a prediction of
any method which can provide the quark condensates $\langle {\bar q}q\rangle(T)$
($q=u,d,s$). Thus, one can get the $T$-dependence of the axion mass (\ref{axionMass})
through the mechanism where DChSB drives the $U_A(1)$ symmetry breaking. (And conversely,
of course: the chiral restoration then drives the restoration of $U_A(1)$ symmetry.)

An excellent tool to study DChSB, and in fact ``produce'' it in the theoretical
sense, is one of the basic equations of the DSE approach - the gap equation.
The most interesting thing it does for nonperturbative QCD is explaining the
notion of the constituent quark mass around $\frac{1}{3}$ of the nucleon mass
$M_N$ by generating them via DChSB,
{\it in the same process which produces the $q\bar q$ condensates.}
 Thanks to this, 
$q\bar q$  condensates can be evaluated from dressed quark propagators.
 Specifically, hadronic-scale large ($\sim M_N/3$ at small momenta $p$) dressed
 quark-mass functions $M_q(p^2) \equiv B_q(p^2)/A_q(p^2)$ are generated
 despite two orders of magnitude lighter current quark masses $m_q$,
 and in fact even in the chiral limit, when $m_q=0$!
 This happens in low-energy QCD thanks to {\it nonperturbative dressing}
 via strong dynamics, making strongly dressed quark propagators $S_q(p)$
 out of the free quark propagators $S^{\scriptstyle {\rm free}}_{q}\,$:
\begin{equation}
S^{\scriptstyle {\rm free}}_{q}(p)=\frac{ 1 }{i\gamma\cdot p + { m}_q}
\,\, \longrightarrow \,\,  S_q(p)=\frac{ 1 }{i \gamma\cdot p \, A_q(p^2) + B_q(p^2)}
\quad \quad \text{(Euclidean space expressions).}
\label{dressingS}
\end{equation}
The solution for the dressed quark propagator $S_q(p)$ of the flavor $q$,
{\it i.e.}, the dressing functions $A_q(p^2)$ and $B_q(p^2)$, are found
by solving the gap equation
\begin{equation}
{S}_q^{-1}(p) \, = \, S^{\scriptstyle {\rm free}}_{q}(p)^{-1} \, - \, \Sigma_q(p)
 \, , \qquad (q=u,d,s) \, ,
\label{gapDSE}
\end{equation}
where $\Sigma_q(p)$ is the corresponding DChSB-generated self-energy,
for example, Equation (\ref{quarkSelf-energy}) if the rainbow-ladder truncation
 is adopted.

In the present work, all we want to model of nonperturbative QCD are 
the condensates $\langle{\bar q}q\rangle$ at all temperatures $T$,
and for that the solutions of the quark-propagator gap Equation
(\ref{gapDSE}) are sufficient, i.e., we do not need the Bethe–Salpeter
 equation (BSE) for the $q{\bar q}'$ pseudoscalar bound states.
 However, we want the same condensates, and basically the same
 (leading term of) the topological susceptibility as we had in our
 related $\eta$-$\eta'$ paper \cite{Horvatic:2018ztu}, and in a number
 of earlier papers, such as Refs. \cite{Horvatic:2007qs,Benic:2011fv}.
 Thus, we now use the same model interaction we have been using then 
 in the consistent rainbow-ladder (RL) truncation of DSE's to produce
 chirally correctly behaving pseudoscalar mesons - that is, with the
 non-anomalous parts of their masses given by GMOR (\ref{GMOR}).
  
Thus, the quark self-energy in the gap Equation (\ref{gapDSE})
 in the RL truncation is
\begin{eqnarray}
 \label{quarkSelf-energy}
\!\!\!\! \Sigma_q(p) =  
- \int \!\!\frac{d^4\ell}{(2\pi)^4} \,
  g^2 D_{\mu\nu}^{ab}(p-\ell)_{\mbox{\rm\scriptsize eff}} \, 
\frac{\lambda^a}{2}\,\gamma^{\mu}
S_q(\ell) \frac{\lambda^b}{2}\,\gamma^{\nu}  , 
\end{eqnarray}
where $D_{\mu\nu}^{ab}(k)_{\mbox{\rm\scriptsize eff}}$ is an 
effective gluon propagator, which should be chosen to model the nonperturbative,
 low-energy domain of QCD. This can be done in varying degrees of DSE modeling,
 depending on the variety of problems one wants to treat
\cite{Alkofer:2000wg,Roberts:2000aa,Holl:2006ni,Fischer:2006ub,Blaschke:2000gd}.
 For example, in the context of low-energy meson phenomenology, if one does not
 aim to address problems of perturbative QCD, it is better {\it not to include}
 the perturbative part of the QCD interaction. Otherwise, in the words of very
 authoritative DSE practitioners, ``the logarithmic tail and its associated
 renormalization represent an unnecessary obfuscation." \cite{Alkofer:2002bp}

In medium, the original $O(4)$ symmetry is broken to $O(3)$ symmetry. The most general
form of the dressed quark propagator then has four independent tensor structures and
four corresponding dressing functions. At nonvanishing temperature, $T > 0$,
we use the Matsubara formalism, where four-momenta decompose into three-momenta
and Matsubara frequencies: $p = (p^0, {\vec p}) \to p_{n} = (\omega_{n}, {\vec p})$.
Therefore, the (inverted) dressed quark propagator ${S}_q(p)$ (\ref{gapDSE})
 becomes 
\begin{equation}
{S}_q^{-1}(p_{n})  = 
S_q^{-1}({\vec{p}}, \omega_n) = i{\vec\gamma\cdot\vec{p}}\, A_q({\vec{p}}^2, \omega_n) 
 + i\, \omega_n \, \gamma_4 \, C_q({\vec{p}}^2, \omega_n) + B_q({\vec{p}}^2, \omega_n)
+ i\, \omega_n  \,\gamma_4 \,{\vec\gamma\cdot{\vec{p}}}\,{\cal D}_q({\vec{p}}^2, \omega_n) .
 \label{SatT}
\end{equation}
(The $T$-dependence of the propagator dressing functions is understood 
and, to save space, is not indicated explicitly, except in the Appendix.)

Nevertheless, the last dressing function ${\cal D}_q({\vec{p}}^2, \omega_n)$
 is so very small that it is quite safe and customary to neglect it
 -- {\it e.g.}, see Refs. \cite{Roberts:2000aa,Contant:2017gtz}.
Thus, also we set ${\cal D}_q \equiv 0$, leaving only $A_q, C_q$ and $B_q$.

For applications in involved contexts, such as calculations at $T>0$,
 appropriate simplifications are very welcome for tractability.
This is why in Refs. \cite{Horvatic:2007qs,Benic:2011fv,Horvatic:2018ztu} and presently,
 we adopted relatively simple, but phenomenologically successful
\cite{Blaschke:2000gd,Horvatic:2007qs,Blaschke:2006ss,Horvatic:2007wu,Horvatic:2007mi}
 separable approximation \cite{Blaschke:2000gd}.
The details on the functional form and parameters of the presently used model
 interaction can be found in the Appendix.

As already pointed out in the original Ref. \cite{Blaschke:2000gd}, the model
 {\it Ans\"atze} for the nonperturbative low-energy interaction (``interaction
form factors'') are such that they provide sufficient ultraviolet suppression.
 Therefore, as noted already in Ref. \cite{Blaschke:2000gd}, no renormalization
  is needed and the
multiplicative renormalization constants, which would otherwise be needed
 in the gap Equation (\ref{gapDSE}) with (\ref{quarkSelf-energy}), are 1.
%
The usual expression for the condensate of the flavor $q$ then becomes
\begin{equation}
\langle \bar{q}q\rangle \, = \, 
- \, N_c \,{\sumint}_p \mbox{\rm Tr} \left[ S_q(p) \right] \, \equiv \, - \, N_c \, T \, 
\sum_{n \in \mathbb{Z}}  \int\, \frac{d^3{p}}{(2\pi)^3} \, \mbox{\rm Tr}
 \left[ S_q({\vec{p}}, \omega_n) \right]\, , \label{qCondensate}
\end{equation}
where $\mbox{\rm Tr}$ is the trace in Dirac space, and the combined integral-sum symbol
indicates that when the calculation is at $T > 0$, the four-momentum integration decomposes
into the three-momentum integration and summation over fermionic Matsubara frequencies
$\omega_n = (2 n + 1)\pi T, \, n \in \mathbb{Z}$.

\begin{figure}[htb]
\centerline{%
\includegraphics[width=14.2cm]{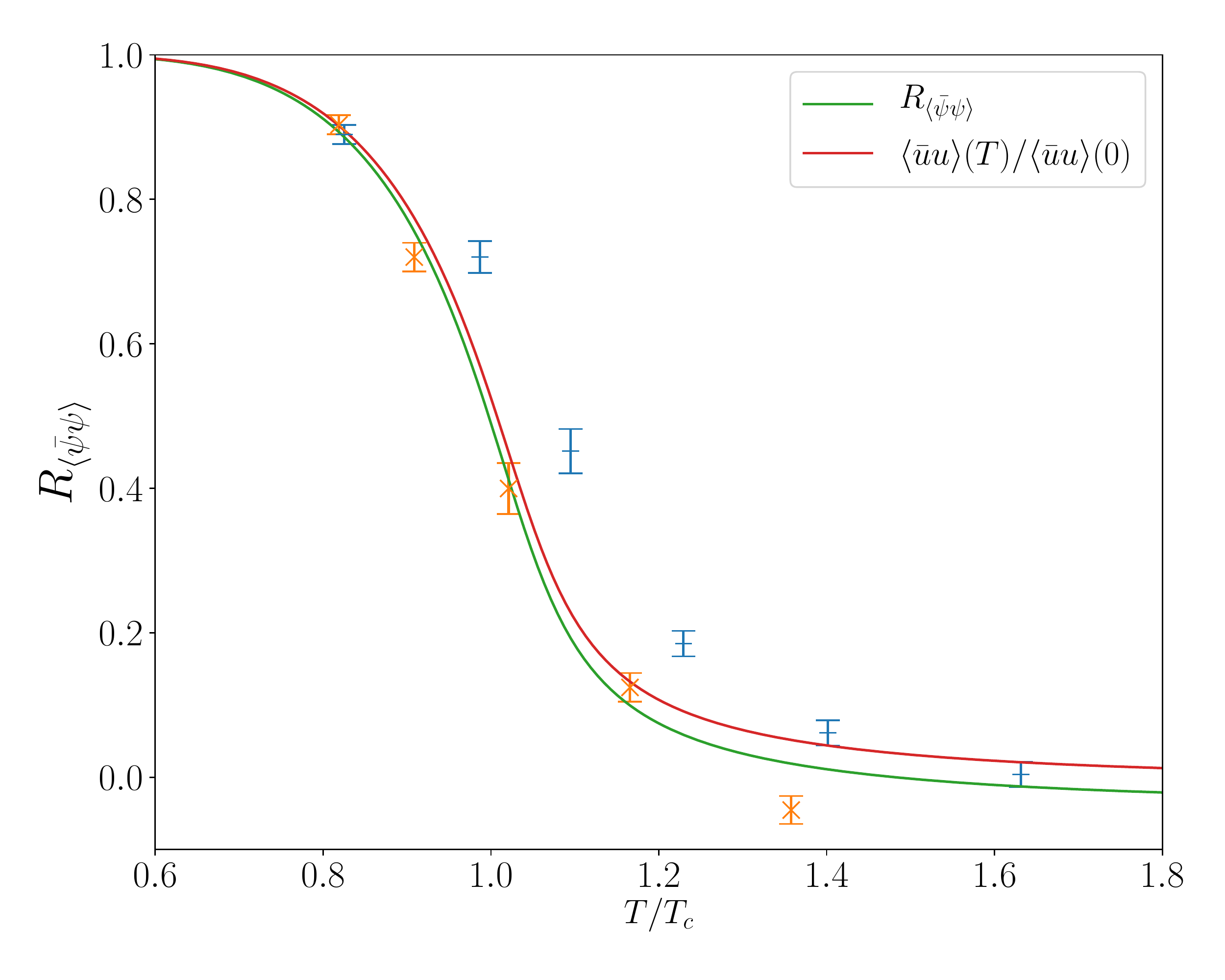}}
\caption{The relative-temperature $T/T_c$ dependence of the subtracted (and normalized) condensate
 $R_{\langle\bar\psi\psi\rangle}$ defined by Equation (\ref{Burger+alPRD87}) and introduced by
Ref. \cite{Burger:2011zc}. The lattice data points are from Figure 6 of Ref. \cite{Kotov:2019dby},
 but scaled for the critical temperatures $T_\chi$ from their Table 2, which is different for
 the ``crosses'' (data points \cite{Kotov:2019dby} for $m_\pi \approx 370$ MeV) and ``bars''
 (data points \cite{Kotov:2019dby} for $m_\pi \approx 210$ MeV). The lower, green curve results
from the $R_{\langle\bar\psi\psi\rangle}$ (\ref{Burger+alPRD87}) subtraction of our $u$-quark
condensate. The upper, red curve is the $T$-dependence of our $u$-quark condensate when
 regularized in the usual way (see text).}
\label{Rpsi_w_cond_rank2}
\end{figure}

As is well known, the condensates (\ref{qCondensate}) are finite only for massless quarks,
$m_q = 0$, i.e., only $\langle \bar{q}q\rangle_0$ is finite, while the ``massive'' 
condensates are badly divergent, and must be regularized, {\it i.e.}, divergences
must be subtracted. Since the subtraction procedure is not uniquely defined, the chiral
condensate at nonvanishing quark mass is also not uniquely defined. However, the arbitrariness
is in practice slight and should rather be classified as fuzziness. It should not be given
too large importance in the light of small differences between the results of various
 sensible procedures.

Our regularization procedure is subtracting the divergence-causing $m_q$ ($\sim$ several MeV)
from the scalar quark-dressing function $B_q(p^2)$ ($\sim$ several hundred MeV) whenever it
 is found in the numerator of the condensate integrand. To justify our particular
 regularization of massive condensates as physically meaningful and sensible, we
 have examined its consistency with two different subtractions used on lattice
\cite{Borsanyi:2010bp,Burger:2011zc,Kotov:2019dby} and in a recent DSE-approach paper \cite{Isserstedt:2019pgx}.

 We shall now test our massive condensates obtained from the separable rank-2 DSE model
 (see the Appendix), whose regularized versions have already been shown in Figure \ref{condens0uds}.

Let us first consider the subtraction on lattice (normalized to 1 for $T=0$) first proposed
in Ref. \cite{Burger:2011zc} in their Equation (17), rewritten in our notation and applied to our
condensate of $u$-quarks (and of course $d$-quarks in the isospin limit):
\begin{equation}
R_{\langle\bar\psi\psi\rangle}(T) \, = \, R_{\langle\bar uu\rangle}(T) \, = \,
\frac{\langle\bar u\,u\rangle(T) - \langle\bar u\,u\rangle(0) + \langle\bar q \, q\rangle_0(0)}
     {\langle\bar q \, q\rangle_0(0)} \, .
\label{Burger+alPRD87}
\end{equation}

In Figure \ref{Rpsi_w_cond_rank2}, the upper, red curve shows (normalized) $u$-quark condensate
$\langle\bar u\,u\rangle(T)/\langle\bar u\,u\rangle(0)$ when regularized in the usual way, by
subtracting $m_q$ from $B_q(p^2)$ in the numerator of the condensate integrand. It agrees very
well with the lattice regularization $R_{\langle\bar uu\rangle}$ (\ref{Burger+alPRD87}) of our
condensate $\langle\bar u\,u\rangle(T)$, represented by the green curve. The agreement with
the lattice data points taken (if pertinent) from Table 6 of Ref. \cite{Kotov:2019dby} is also
 rather good.

\begin{figure}[htb]
\centerline{%
\includegraphics[width=14.2cm]{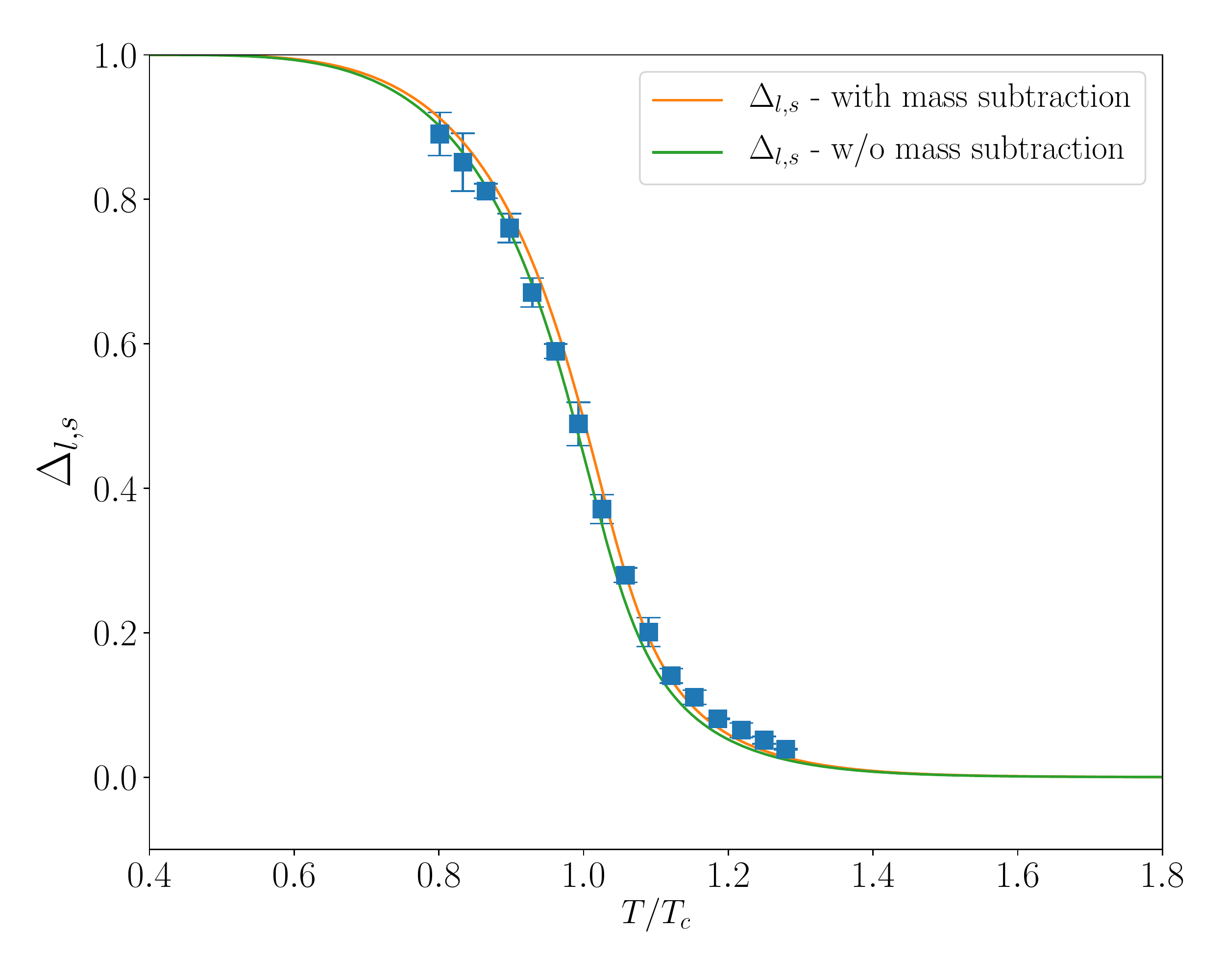}}
\caption{The relative-temperature $T/T_c$ dependence of the (normalized) subtracted quark condensate
(\ref{normalizLattSubtract}) from the lattice \cite{Borsanyi:2010bp} (blue squares) and from our
condensates. Slightly lower, green curve results from our unsubtracted condensates plugged in Equation
 (\ref{normalizLattSubtract}), while the very slightly higher, red curve is from our already
 subtracted condensates.}
\label{Delta_us_rank2}
\end{figure}

Next, we examine the consistency of our subtraction with the most usual
 condensate subtraction on the lattice, which combines the light and strange
 quark condensates and their masses like this:
\begin{equation}
\bar\Delta_{{\scriptsize l,s}}(T) \, = \, \langle\bar l \, l\rangle_{{\scriptsize l}}(T)
 \, - \, \frac{m_{{\scriptsize l}}}{m_{{\scriptsize s}}} \,
\langle\bar s \, s\rangle_{{\scriptsize s}}(T) \, .
\end{equation}
Following Isserstedt {\it et al.} \cite{Isserstedt:2019pgx}, in Figure \ref{Delta_us_rank2}
   we make comparison of the normalized version thereof
\begin{equation}
\Delta_{{\scriptsize l,s}}(T) \,  = \, 
\frac{\,\,\langle\bar l\, l\rangle(T) \, - \, \frac{m_{\scriptsize l}}{m_{\scriptsize s}}\, \langle\bar s \, s\rangle(T) \,\,}{\langle\bar l\, l\rangle(0) \, - \, \frac{m_{\scriptsize l}}{m_{\scriptsize s}} \langle\bar s \, s\rangle(0)}
\qquad  {\mbox{($l\, = \,u\,$ or $\,d\,$ in the isospin symmetric limit) }}
\label{normalizLattSubtract}
\end{equation}
with the lattice data of Ref. \cite{Borsanyi:2010bp}. The agreement is very good, which implies
also the agreement with the subtracted condensates in the recent DSE paper \cite{Isserstedt:2019pgx},
which made this successful comparison first (in its Figure 3).

To conclude: results shown in Figures \ref{Rpsi_w_cond_rank2}  and \ref{Delta_us_rank2}
demonstrate that certain arbitrariness in the choice of regularization does not disqualify
our massive condensates from useful applications, such as using them in Equation (\ref{chiLIGHTq})
to make predictions on the topological susceptibility.

 \vskip 5mm

\subsection{Axion Mass and Topological Susceptibility---Results from the Rank-2 Separable Model in the Isosymmetric Limit}

\vskip 2mm

\begin{figure}[htb]
\centerline{%
\includegraphics[width=14.2cm]{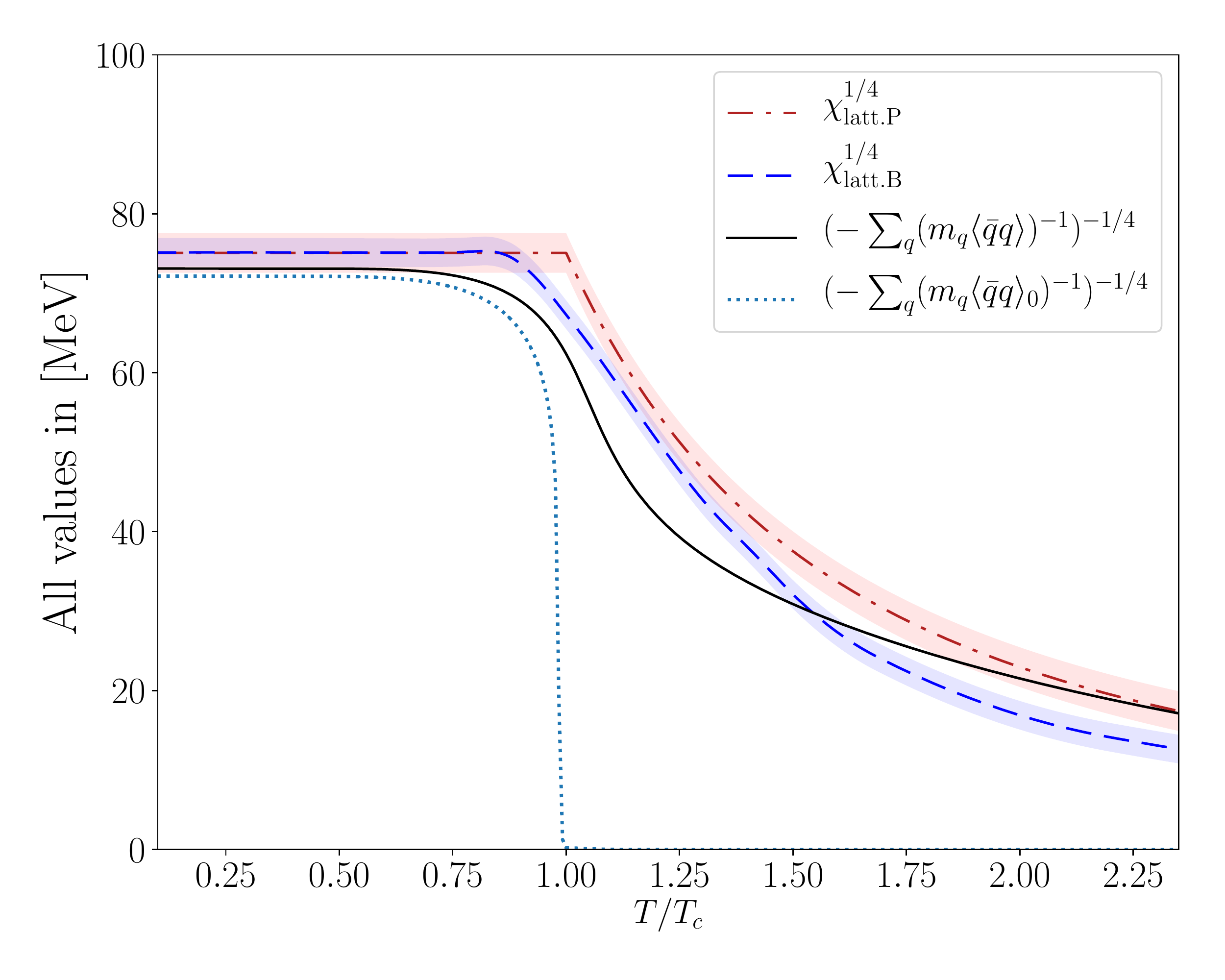}}
\caption{The relative-temperature $T/T_c$ dependence of  
   (the leading term of) $\, \chi(T)^{1/4} \,$ from our often adopted 
\cite{Horvatic:2007qs,Horvatic:2007wu,Horvatic:2007mi,Benic:2011fv,Horvatic:2018ztu}
isosymmetric DSE rank-2 separable model: solid curve for Equation (\ref{chiLIGHTq})
 with massive-quark condensates, while the dotted curve results from using
 $\langle{\bar q}q\rangle_0$ instead.
 $\, \chi(T)^{1/4} \,$ (with uncertainties) from lattice: dash-dotted curve extracted
 from Petreczky {\it et al.,} \cite{Petreczky:2016vrs} and long-dashed curve, 
 from Borsany {\it et al.,} \cite{Borsanyi:2016ksw}.  (Colors online.)}
\label{FigIsosymm}
\end{figure}

Our result for $\chi(T)^{1/4}=\sqrt{m_{\rm a}(T) \, f_{\rm a}}$ is presented in
Figure \ref{FigIsosymm} as a solid curve and compared, up to $T\approx 2.3\, T_c$,
with the corresponding results of two lattice groups
\cite{Petreczky:2016vrs,Borsanyi:2016ksw}, rescaled to the relative-temperature $T/T_c$.
{(Table \ref{ourTable} gives numerical values of our results at $T=0$.)}

In our case, the results for $\chi(T)$ and condensates $\langle{\bar u}u\rangle(T)$,
$\langle{\bar d}d\rangle(T)$ and $\langle{\bar s}s\rangle(T)$ needed to obtain it, are
 predictions of the dynamical DSE model used in the $T>0$ study of $\eta'$-$\eta$
\cite{Horvatic:2018ztu}. This is the same modeling of the low-energy, nonperturbative QCD
interactions as we have already employed in our earlier studies of light pseudoscalar mesons
at $T \geq 0$ \cite{Horvatic:2007qs,Horvatic:2007wu,Horvatic:2007mi,Benic:2011fv}: the
separable model interaction - see, {\it e.g.,} \cite{Cahill:1998rs,Blaschke:2000gd}, and
references therein. We have adopted the so-called rank-2 variant from Ref. \cite{Blaschke:2000gd}.
The adopted model with our choice of parameters is defined in detail in the Appendix of the
present work, after the subsection II.A of Ref. \cite{Horvatic:2007qs} and Ref. \cite{Blaschke:2007ce}.
It employs the model current-quark-mass parameters $m_u = m_d \equiv m_l = 5.49$ MeV
and ${m}_s = 115$ MeV. The model prediction for condensates at $T=0$ are
$\langle{\bar s}s\rangle=(-238.81\;\rm MeV)^3$ for the heaviest quark, while
 isosymmetric condensates of the lightest flavors, $\langle{\bar u}u\rangle =
\langle{\bar d}d\,\rangle\equiv\langle\,{\bar l}\,l\rangle=(-218.69\;\rm MeV)^3$ are
quite close to the ``massless'' one, $\langle{\bar q}q\rangle_0 = (-216.25\;\rm MeV)^3$.

Contrary to, {\it e.g.}, Ref. \cite{Benic:2011fv}, where the condensate of massless
quarks $\langle{\bar q}q\rangle_0(T)$ was used, in Ref. \cite{Horvatic:2018ztu} and
here we follow Shore \cite{Shore:2006mm} in using condensates of light quarks with
nonvanishing current masses. The smooth, crossover behavior around the pseudocritical
temperature $T_c$ for the chiral transition (now confirmed at vanishing baryon density
 by lattice studies such as 
\cite{Aoki:2012yj,Buchoff:2013nra,Dick:2015twa,Petreczky:2016vrs,Borsanyi:2016ksw}),
is obtained thanks to the DChSB condensates of realistically massive light quarks
 -- {\it i.e.,} the quarks with realistic explicit chiral symmetry breaking
 \cite{Horvatic:2018ztu}.

In contrast, using in Equation (\ref{chiLIGHTq}) the massless-quark condensate
$\langle {\bar q}q \rangle_0$ (which drops sharply to zero at $T_c$)
instead of the ``massive'' ones,
would dictate a sharp transition of the second order at $T_c$
\cite{Benic:2011fv,Horvatic:2018ztu} also for $\chi(T)$, illustrated 
in Figure \ref{FigIsosymm} by the dotted curve. This would of course imply
that axions are massless \cite{Klabucar:SchladmingAPP} for $T > T_c$.
It is of academic interest to know what consequences would be
thereof for cosmology, but now it is clear that only crossover
is realistic \cite{Petreczky:2016vrs,Borsanyi:2016ksw}.

The rather good agreement with lattice in Figure \ref{FigIsosymm} resulted without
 any refitting of this model, either in Ref. \cite{Horvatic:2018ztu}
 for $\eta'$ and $\eta$, or in this subsection. The model is in the isosymmetric
limit, ${m}_u = {m}_{d} \equiv {m}_{l}$, which is perfectly adequate for
most purposes in hadronic physics. Nevertheless, the QCD topological susceptibility
 $\chi$ in its version (\ref{chiLIGHTq}) contains the current quark masses in the
 form of harmonic averages of $m_q\, \langle {\bar q}q\rangle$ ($q=u,d,s$).
 A harmonic average is dominated by its smallest argument, and presently this is
 the lightest current-quark-mass parameter, motivating us to investigate the changes
occurring beyond the isospin symmetric point.

\begin{table}
\centering
\begin{tabular}{|c|cccccccc|}
\hline  
 &&&&&&&& \\
$T = 0$   &      $m_u$ & $m_d$ & $m_s$ &$\chi_0$ & $\langle \bar{u}u\rangle$ & $\langle \bar{d}d\rangle$ & $\langle \bar{s}s\rangle$ & $\chi$  \\
  &&&& (with $\langle q\bar{q}\rangle_0$) &&&& \\
\hline
  rank-2 &&&&&&&& \\
\hline
        &&&&&&&&  \\
$m_u=m_d$ \cite{Horvatic:2018ztu} & 5.49 &  5.49 & 115 & $72.18^4$ &  $-218.69^3$ &  $-218.69^3$ &  $-238.81^3$ &   $72.73^4$ \\
 &&&&&&&&  \\
with constraint  &&&&&&&&  \\
$m_u=0.48\, m_d$ \cite{Tanabashi:2018oca}, &  4.66 &  9.71 & 115 & $74.64^4$ &     $-218.35^3$ &     $-220.33^3$ &  $-238.81^3$ &     $75.44^4$ \\
fitted $m_u$ \& $m_d$ &&&&&&&&  \\
\hline
  rank-1 &&&&&&&& \\
\hline
&&&&&&&&  \\
$m_u=m_d$ &     6.6 &   6.6 &   142 & $83.87^4$ &  $-249.27^3$ &    $-249.27^3$ &                  $-251.49^3$ &           $84.08^4$ \\
 &&&&&&&&  \\
with constraint &&&&&&&&  \\
$m_u=0.48\, m_d$ \cite{Tanabashi:2018oca}, &  3.15 &  6.56 &  142 & $75.31^4$ &   $-248.87^3$ & $-249.21^3$ &     $-251.49^3$  & $75.43^4$ \\
fitted $m_u$ \& $m_d$ &&&&&&&&  \\
\hline
\end{tabular}
              \\
\caption{For the both variants of the DSE separable model (with the rank-2 and rank-1 interaction Ansatz) used in the
present paper, various sets of values of the model quark-mass parameters $m_q$ $(q=u, d, s)$ are related to the model
results for the topological susceptibility $\chi$ and the ``massive" condensates $\langle \bar{q}\, q\rangle$ at $T=0$. 
The topological susceptibility $\chi$ varies because of varying $m_q$ and (to a much lesser extent) because of the
changes of ``massive" condensates induced by changes of the quark-mass parameters $m_q$. The massless-quark condensate,
$\langle \bar{q}\, q\rangle_0$, depends only on the dynamical DSE model: always $\langle \bar{q}\, q\rangle_0 = -216.25^3$ MeV$^3$
for the rank-2 model, and $\langle \bar{q}\, q\rangle_0 = -248.47^3$ MeV$^3$ for the rank-1 model. Thus, the topological
susceptibility $\chi_0$ calculated with the chiral-limit condensate, varies for a given model only because of varying
values of $m_q$.  All values are in MeV (or the indicated 3rd or 4th powers of MeV). }
\label{ourTable}
\end{table}

\vskip 2mm

\subsection{Axion Mass and Topological Susceptibility from Rank-1 and Rank-2 Models out of the Isosymmetric Limit}

The previous isosymmetric case, pertinent also for the $\eta'$-$\eta$ study \cite{Horvatic:2018ztu},
 has the current-quark-mass model parameters ${m}_u = {m}_{d} = 5.49$ MeV. This
 is above the most recent PDG quark-mass values \cite{Tanabashi:2018oca}, but
 anyway yields $\chi(T=0)=(72.73 \;\rm MeV)^4$ already a little below the lattice
 results \cite{Petreczky:2016vrs,Borsanyi:2016ksw}, and below the most recent
 chiral perturbation theory result $\chi(T=0) = (75.44\;\rm MeV)^4$ \cite{Gorghetto:2018ocs}.

This seems not to bode well for the attempts out of the isosymmetric limit, because
 lowering the values of the current masses seems to threaten yielding unacceptably
 low values of the topological susceptibility. Indeed, taking the central values
from the current quark masses $m_u=2.2^{+0.5}_{-0.4}$ MeV and $m_d=4.70^{+0.5}_{-0.3}$ MeV
 and $m_s=95^{+9}_{-3}$ MeV recently quoted by PDG \cite{Tanabashi:2018oca}, yields
 just $(62.50 \;\rm MeV)^4$ for the leading term of Equation (\ref{chiLIGHTq}) at $T=0$.

However, our model $m_u, m_d$ and $m_s$ are phenomenological current-quark-mass
 {\it parameters}, and cannot be quite unambiguously and precisely related to
 the somewhat lower PDG values of the current quark masses. The better relation
 is through the {\it ratios} of quark masses, for which PDG gives 
 $m_u/m_d = 0.48^{+0.07}_{-0.08}\;$ \cite{Tanabashi:2018oca}.

We thus require that $m_u^{\rm fit}/m_d^{\rm fit} = 0.48$  be satisfied by the
 new non-isosymmetric mass parameters $m_u^{\rm fit}$ and $m_d^{\rm fit}$ when
 they are varied to reproduce the recent most precise value
 $\chi(T=0) = (75.44\;\rm MeV)^4$ \cite{Gorghetto:2018ocs}.
 We get ${m}_u^{\rm fit} = 4.66$ MeV, resulting in the condensate
 $\langle{\bar u}u\rangle(T=0) = (-218.35\;\rm MeV)^3$ and $m_d^{\rm fit} = 9.71$ MeV,
 resulting in $\langle{\bar d}d\rangle(T=0) = (-220.33\;\rm MeV)^3$.
 (The $s$-mass parameter is not varied, {\it i.e.,} $m_s \equiv m_s^{\rm fit}$. The rest
 of model parameters, namely those in the {\it Ansatz} functions ${\cal F}_0(p^2)$ and ${\cal F}_1(p^2)$
 modeling the strength of the rank-2 nonperturbative interaction (see the Appendix), are also not varied.)

\begin{figure}[htb]
\centerline{%
\includegraphics[width=14.8cm]{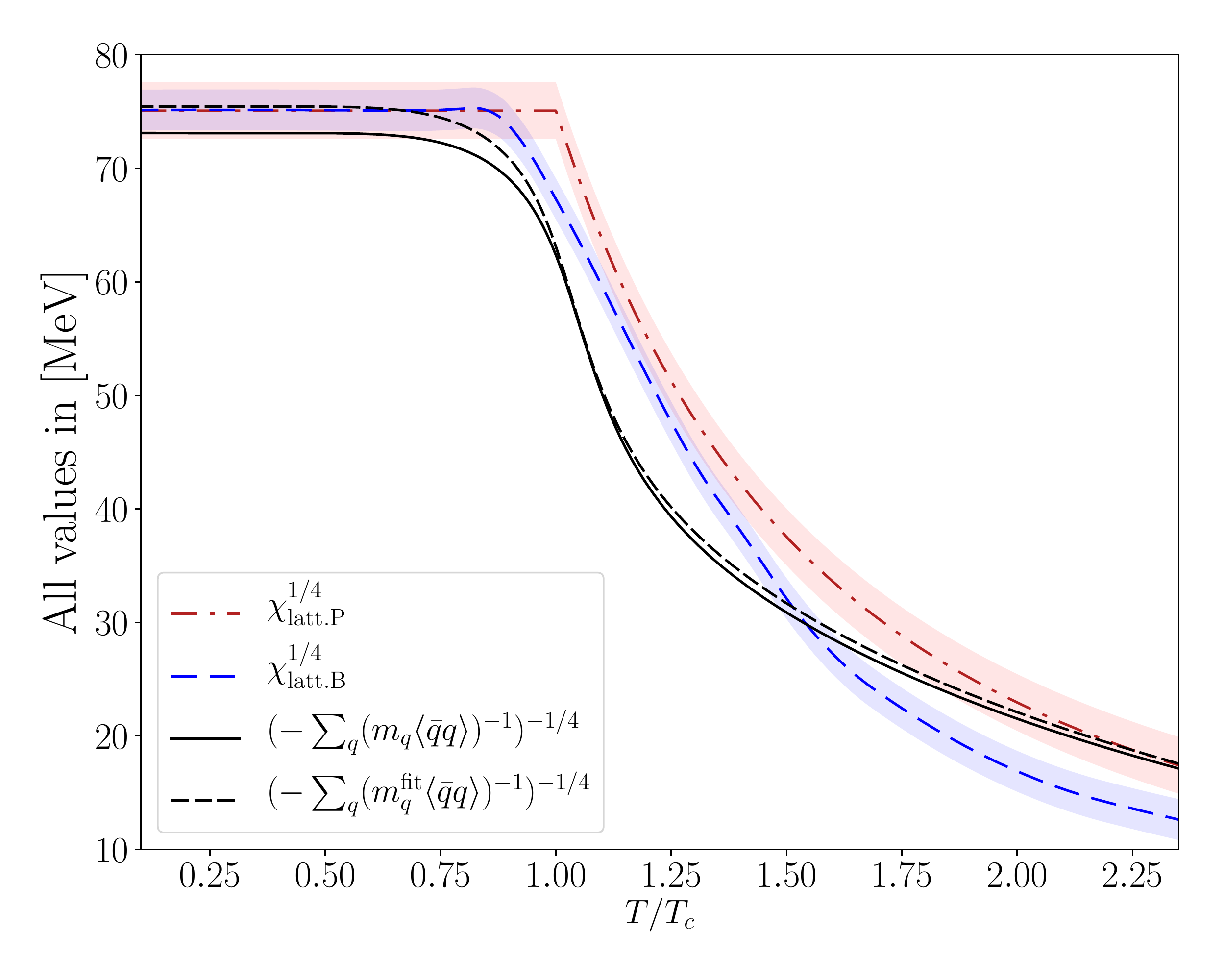}}
\caption{The short-dashed black curve shows the non-isosymmetric case of the leading
term of $\,\chi(T)^{1/4}\,$, Equation (\ref{chiLIGHTq}), with $\,\, m_u^{\rm fit} = 4.66\,$
MeV and $\, m_d^{\rm fit} = m_u^{\rm fit}/0.48 = 9.71\,$ MeV, and appropriately
recalculated condensates $\langle{\bar u}u\rangle(T)$ and $\langle{\bar d}d\rangle(T)$.
The vertical scale is zoomed with respect to Figure \ref{FigIsosymm} to help resolve the
short-dashed curve from the solid curve representing again the isosymmetric case of
the same separable rank-2 model. Also, the lattice results
\cite{Petreczky:2016vrs,Borsanyi:2016ksw} are again depicted as in Figure \ref{FigIsosymm}.
}
\label{FigRank2all}
\end{figure}

 The $T$-dependence of the resulting $\chi(T)^{1/4}$ is given by the short-dashed
 black curve in Figure \ref{FigRank2all}. Except its better agreement with the lattice
 results \cite{Petreczky:2016vrs,Borsanyi:2016ksw} at low $T$, the new (dashed)
 $\chi(T)^{1/4}$ curve is very close to the isosymmetric (solid) curve.

\begin{figure}[htb]
\centerline{%
\includegraphics[width=14.2cm]{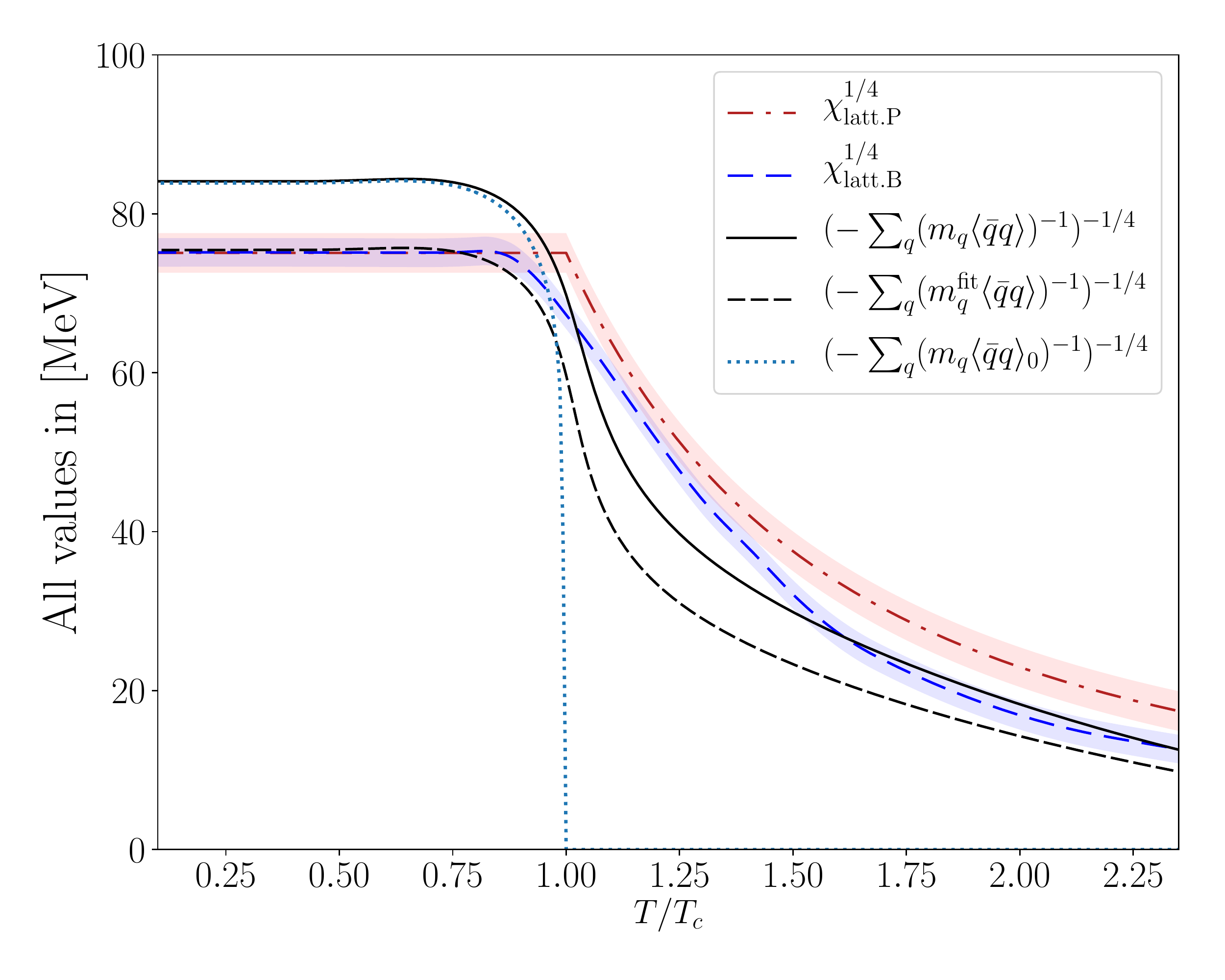}}
\caption{From the calculation in the separable rank-1 DSE model \cite{Blaschke:2000gd},
the relative-temperature $T/T_c$ dependence of (the leading term of) $\,\chi(T)^{1/4}\,$
is represented by:  {\it (i)} the solid curve for the isosymmetric case with
$\, m_u = m_d = 6.6$ MeV and $\, m_s = 142$ MeV, {\it (ii)} the dotted curve
is for the same mass parameters, but with all condensates approximated by the
 ``massless'' condensate $\langle{\bar q}q\rangle_0(T)$, and {\it (iii)} the
 short-dashed curve for the non-isosymmetric case $\, m_u^{\rm fit}=3.15$ MeV,
$\, m_d^{\rm fit} = 6.56$ MeV, while $\, m_s^{\rm fit} \equiv m_s = 142$ MeV.
 The pertinent lattice results are presented in the same way as in the previous
 two Figures: the dash-dotted and long-dashed curves extracted, respectively,
 from Refs. \cite{Petreczky:2016vrs} and \cite{Borsanyi:2016ksw}.  Colors online.}
\label{FigRank1all}
\end{figure}

Now we will check the model dependence by comparing our results
presented so far (obtained in the rank-2 model) with those we 
get in the separable rank-1 model of Ref. \cite{Blaschke:2000gd}.
It is similar to the previously considered rank-2 one by modeling the
low-energy, nonperturbative QCD interaction with an {\it Ansatz}
separating the momenta $p_a, p_b$ of interacting constituents, but
is of a simpler form, proportional to just ${\cal F}_0(p_a^2)\, {\cal F}_0(p_b^2)$.
Its presently interesting feature is that for similar quark-mass
parameters, it yields significantly larger condensates than those
 in the separable rank-2 model, and thus also larger $\chi$.
(This also holds at low and vanishing $T$ even for $\chi(T)$ calculated
using only the ``massless'' condensate $\langle{\bar q}q\rangle_0(T)$.
This case is depicted in Figure \ref{FigRank1all} as the dotted curve.) 

The original rank-1 model employs the light-quark current mass parameters
in the isosymmetric limit: $m_u=m_d \equiv m_l=6.6$ MeV \cite{Blaschke:2000gd}.
However, in Equation (\ref{chiLIGHTq}) we also need the $s$-flavor. The fit to the kaon
mass yields $m_s = 142$ MeV. The model prediction for condensates at $T=0$ are
then $\langle{\bar s}s\rangle=(-251.49\;\rm MeV)^3$ for the heaviest quark, while
 isosymmetric condensates of the lightest flavors, $\langle{\bar u}u\rangle =
\langle{\bar d}d\,\rangle\equiv\langle\,{\bar l}\,l\rangle=(-249.27\;\rm MeV)^3$ are
quite close to the ``massless'' one, $\langle{\bar q}q\rangle_0 = (-248.47\; \rm MeV)^3$.
 This gives too large topological susceptibility at $T=0$, namely
 $\chi(0)= (84.08 \;\rm MeV)^4$.  Nevertheless, for large $T$,
 it also falls with $T$ somewhat faster than the rank-2 $\chi(T)$,
 since rank-1 condensates fall with $T$ somewhat faster than the rank-2 ones.

The isosymmetric rank-1 $\chi(T)$ is depicted by the solid black curve in
 Figure \ref{FigRank1all},
showing that it actually falls with $T$ faster even than $\chi(T)$'s from lattice
\cite{Petreczky:2016vrs,Borsanyi:2016ksw} for practically all $T$'s high enough to
 induce changes.
Then, comparing Figures \ref{FigRank2all}  and \ref{FigRank1all} shows that the lattice
high-$T$ results are in between high-$T$ results of the two separable models.

To go out of the isospin limit with the rank-1 model, we again require that the
changed parameters $m_u^{\rm fit}$ and $m_d^{\rm fit}$ (together with the condensates
resulting from them) fit the currently most precise $T=0$ value of the topological
susceptibility, $\chi = (75.44\;\rm MeV)^4$  \cite{Gorghetto:2018ocs}, while also
obeying $m_u^{\rm fit}/m_d^{\rm fit} = 0.48$, {\it i.e.,} the central value of
 the PDG \cite{Tanabashi:2018oca} $m_u/m_d$ ratio.  (Again, other
 model parameters including $m_s$ are not varied: $m_s \equiv m_s^{\rm fit}$.)

In our rank-1 model, these requirements yield $m_d^{\rm fit}=6.56$ MeV, {\it i.e.,} it
practically remained the same as in the originally fitted model \cite{Blaschke:2000gd}.
Of course, its condensate $\langle{\bar d}d\rangle$ also remains the same.
The lightest flavor has the mass parameter lowered to $m_u^{\rm fit} = 3.15$ MeV.
Now it has only slightly lower condensate $\langle{\bar u}u\rangle(T=0) = (-248.87\;\rm MeV)^3$,
which is even closer to the ``massless'' $\langle{\bar q}q\rangle_0(T=0)$. Nevertheless,
$\langle{\bar u}u\rangle(T)$ retains the crossover behavior for $T>0$, although
 it falls with $T$ steeper than more ``massive'' condensates.

The resulting non-isosymmetric $\chi(T)^{1/4} =\sqrt{m_{\rm a}(T) \, f_{\rm a}}$ is in
 Figure \ref{FigRank1all} shown as the short-dashed black curve, which is everywhere
 consistently the lowest (among the ``massive'', crossover curves).

\section{Summary and Discussion}

In the DSE framework, we have obtained predictions for the nontrivial part of
 the $T$-dependence of the axion mass $m_{\rm a}(T) = \sqrt{\chi(T)}/f_{\rm a}$,
 Equation (\ref{axionMass}),
 by calculating the QCD topological susceptibility $\chi(T)$, since the
 unknown Peccei–Quinn scale $f_{\rm a}$ is just an overall constant. 
We have used two empirically successful dynamical models of the separable type
\cite{Blaschke:2000gd} to model nonperturbative QCD at $T > 0$. We also studied
the effects of varying the mass parameters of the lightest flavors
 out of the isospin limit, and found that our $\chi(T)$, and consequently
$m_{\rm a}(T)$, are robust with respect to the non-isosymmetric refitting of
 $\, m_u\,$ and $\, m_d =  m_u/0.48\,$.

All these results of ours on $\chi(T)$, and consequently the related axion mass,
 are in satisfactory agreement with the pertinent lattice results
 \cite{Petreczky:2016vrs,Borsanyi:2016ksw}, and in qualitative
 agreement with those obtained in the NJL model \cite{Lu:2018ukl}. 
 Everyone obtains qualitatively similar crossover of $\chi(T)$ around $T_c$,
 but it would be interesting to speculate what consequences for cosmology
 could be if $\chi(T)$, and thus also $m_{\rm a}(T)$, would abruptly
 fall to zero at $T=T_c$ due to a sharp phase transition of the ``massless''
 condensate $\langle{\bar q}q\rangle_0(T)$.
 Of course, dynamical models of QCD can access only much smaller
 range of temperatures than lattice, where $T \sim 20\, T_c$ has already
 been reached \cite{Borsanyi:2016ksw}. 
 (On the other hand, the thermal behavior of the $U_A(1)$ anomaly
 could not be accessed in chiral perturbation theory \cite{Gu:2018swy}.)

Since it is now established that (at vanishing and low density) the chiral transition
is a crossover, it is important that one can use massive-quark condensates, which
 exhibit crossover behavior around $T\sim T_c$. In the present work, they give us
 directly, through Equation (\ref{chiLIGHTq}), the crossover behavior of $\chi(T)$.
 However, these are regularized condensates, because a nonvanishing current
 quark mass $m_q$ makes the condensate $\langle{\bar q}q\rangle$ plagued by
 divergences, which must be subtracted. In Section 3, we have shown that our
 regularization procedure is reasonable and in good agreement with at
 least two widely used subtractions on the lattice.

To discuss our approach from a broader perspective, it is useful to recall that
JLQCD collaboration \cite{Fukaya:2017wfq} has recently pointed out how the chiral
symmetry breaking and $U_A(1)$ anomaly are tied, and stressed the importance
of the $q\bar q$ chiral condensate in that. The axion mass presently
provides a simple example thereof: through Equation (\ref{axionMass}),
 $m_{\rm a}(T)$ is at all temperatures  directly  expressed   by the
QCD topological susceptibility $\chi(T)$, which is a measure of $U_A(1)$ breaking
by the axial anomaly. We calculate $\chi(T)$ through Equation (\ref{chiLIGHTq}) from
the quark condensates, which in turn arise from DChSB. In addition, conversely: melting of
 condensates around $T\sim T_c$ signals the restoration of the chiral symmetry. 
Therefore, the $U_A(1)$ symmetry breaking and restoration being driven by the chiral
ones is straightforward.

 The relation of $\chi(T)$ to the $\eta'$ mass is, however, a little less
 straightforward \cite{Horvatic:2018ztu} because it involves several
 other elements, but the topological susceptibility remains the main one.
 Since our present results on the axion are, in a way, a by-product of the
 framework which was initially formulated to understand better the
 $T$-dependence of $\eta'$ and $\eta$ masses, we have explained it in detail
 in Subsection 2.2. Therefore, here in the Summary, we should just stress
 that the topological susceptibility $\chi(T)$ is the strong link between
 the QCD axion and the $\eta$-$\eta'$ complex. It so also in the case of the
 present paper regarding our $\eta$-$\eta'$ reference \cite{Horvatic:2018ztu}: 
 specifically, we should note that the actual $T$-dependence of $\eta'$ and $\eta$ is
 rather sensitive to the behavior of $\chi(T)$, and rather accurate $\chi(T)$
 is needed to get acceptable $M_\eta(T)$ and $M_{\eta'}(T)$. Thanks to
 its crossover behavior, our $\chi(T)$ gives in Ref. \cite{Horvatic:2018ztu}
 empirically allowed $T$-dependence of the masses in the $\eta$-$\eta'$
 complex.  However, even a crossover, if it were too steep, would lead to the
 unwanted (experimentally never seen) drop of the $\eta$ mass, just as a
 too slow one would not yield the drop of the $\eta'$ mass required
 according to some experimental analyses \cite{Csorgo:2009pa,Vertesi:2009wf}.

In that sense, our present predictions on $m_{\rm a}(T)$ are thus supported
 by the fact that our calculated topological susceptibility $\chi(T)$
 gives the $T$-dependence of the $U_A(1)$ anomaly-influenced masses
 of $\eta'$ and $\eta$ mesons \cite{Horvatic:2018ztu} which is
consistent with experimental evidence \cite{Csorgo:2009pa,Vertesi:2009wf}.


%
%
%
%
\vspace{6pt} 



\authorcontributions{Conceptualization, D.Kl.; methodology, D.Kl., D.H. and D.Ke.; software, D.H.; validation, D.Kl., D.H. and D.Ke.; formal analysis, D.H., D.Kl. and D.Ke.; investigation, D.H., D.Kl. and D.Ke.; data curation, D.H.; writing--original draft preparation, D.Kl.; writing--review and editing, D.Kl., D.Ke. and D.H.; visualization, D.H., D.Ke.; supervision, D.Kl.}


\acknowledgments{This work was supported in part by
STSM grants from COST Actions CA15213 THOR and CA16214 PHAROS.}

\conflictsofinterest{The authors declare no conflict of interest.}

\abbreviations{The following abbreviations are used in this manuscript:\\

\noindent 
\begin{tabular}{@{}ll}
DChSB  &  Dynamical chiral symmetry breaking  \\
DSE & Dyson–Schwinger equations    \\
QCD & Quantum chromodynamics \\
ABJ & Adler-Bell-Jackiw \\
CKM & Cabibbo-Kobayashi-Maskawa \\
CP  & charge conjugation parity
\end{tabular}}


\appendixtitles{yes} 
\appendixsections{yes}
\appendix
\section{$\quad$ Separable interaction models for usage at $T\geq 0$}

At $T=0$, the Dyson–Schwinger equation (DSE) approach in the rainbow-ladder
 approximation (RLA) tackles efficiently
 solving Dyson–Schwinger gap equation and Bethe–Salpeter equations,
 but extending this to $T>0$ is technically quite difficult.
We thus adopt a simple model for the strong dynamics from Ref.
\cite{Blaschke:2000gd}, namely the model we already used in Refs.
\cite{Horvatic:2007mi,Horvatic:2007qs,Benic:2011fv,Horvatic:2018ztu}{.}
For the effective gluon propagator in a Feynman-like gauge, we use
the {\it separable Ansatz}:
\begin{equation}
g^2 \, D_{\mu\nu}^{ab}(p-\ell)_{\mbox{\rm\scriptsize eff}} \, = \,
\delta^{ab} \, g^2 \, D_{\mu\nu}^{\mathrm{eff}} (p-\ell) \, \longrightarrow \,
\delta_{\mu\nu} \, D(p^2, \ell^2,p\cdot \ell) \, \delta^{ab} \, ,
\label{Feynmangauge}
\end{equation}
whereby the dressed quark-propagator gap Equation (\ref{gapDSE})
 with (\ref{quarkSelf-energy}) yields
\begin{eqnarray}
\label{gapB}
B_q(p^2) \; =  \;  m_q \, + \,  
\frac{16}{3} \int \! \frac{d^4 \ell}{(2\pi)^4} \, D(p^2,\ell^2, p \cdot \ell)
\,\, \frac{B_q(\ell^2)}{\, \ell^2 \, A_q^2(\ell^2) \, + \, B_q^2(\ell^2) \,}\\
\left[ A_q(p^2) \, - \, 1 \, \right] \, p^2  \; =  \;
\frac{8}{3} \int \! \frac{d^4 \ell }{(2\pi)^4}\, D(p^2, \ell^2,p \cdot \ell)
\,\, \frac{(p\cdot \ell) A_q(\ell^2)}{\, \ell^2 A_q^2(\ell^2)+B_q^2(\ell^2)\,}  \, .
\label{gapA}
\end{eqnarray}
More specifically, the so-called {\it rank-2} separable interaction entails:
\begin{equation}
D(p^2,\ell^2,p\cdot \ell) \, = \, D_0 \, {\cal F}_0(p^2) \, {\cal F}_0(\ell^2)
\, + \, D_1 \, {\cal F}_1(p^2) \, (p\cdot \ell ) \, {\cal F}_1(\ell^2)~.
\label{sepAnsatz}
\end{equation}
Then, the solutions of Equations (\ref{gapB})-(\ref{gapA}) for the dressing
 functions are of the form
\begin{equation}
B_q(p^2) =  m_q + b_q\; {\cal F}_0(p^2)
\qquad {\rm and} \qquad
A_q(p^2)=1+a_q \; {\cal F}_1(p^2),
\label{sepSolutions}
\end{equation}
reducing Equations (\ref{gapB})-(\ref{gapA}) to the nonlinear system of equations
 for the constants $b_q$ and $a_q$:
\begin{eqnarray}
\label{gap-b_q}
b_q \; =  \;  
\frac{16D_0}{3} \int \! \frac{d^4 \ell}{(2\pi)^4} \, 
\,\, \frac{{\cal F}_0(p^2) \, B_q(\ell^2)}{\, \ell^2 \, A_q^2(\ell^2) \, + \, B_q^2(\ell^2) \,}\\
 a_q\; =  \;
\frac{2D_1}{3} \int \! \frac{d^4 \ell }{(2\pi)^4}\,
\,\, \frac{\ell^2{\cal F}_1(\ell^2)\, A_q(\ell^2)}{\, \ell^2 A_q^2(\ell^2)+B_q^2(\ell^2)\,}  \, .
\label{gap-a_q}
\end{eqnarray}
%
If one chooses that the second term in the interaction (\ref{sepAnsatz}) is
vanishing, by simply setting to zero the second strength constant, $D_1 = 0$,
one has a still simpler {\it rank-1 separable Ansatz}, where $A_q(p^2) = 1$.

The analytic properties of these model interactions are defined by the choice
of the interaction ``form factors'' ${\cal F}_{0}(p^2)$ and ${\cal F}_{1}(p^2)$.
In the present work we will use the functions \cite{Blaschke:2007ce,Horvatic:2007wu}
\begin{equation}
{\cal F}_0(p^2) = \exp(-p^2/\Lambda_0^2)
\qquad {\rm and} \qquad
{\cal F}_1(p^2) = \frac{1+\exp(-p_0^2/\Lambda_1^2)}
{1+\exp((p^2-p_0^2)/\Lambda_1^2)}~,
\label{F0and1}
\end{equation}
which satisfy the constraints ${\cal F}_0(0)={\cal F}_1(0)=1$ and
${\cal F}_0(\infty)={\cal F}_1(\infty)=0$.

For the numerical calculations we fix the free parameters of the model at $T=0$
 as in Refs.~\cite{Blaschke:2007ce,Horvatic:2007wu},
to reproduce in particular the vacuum masses of the pseudoscalar and vector mesons,
$M_\pi=140$ MeV, $M_K=495$ MeV, $M_\rho=770$ MeV, the pion decay constant $f_\pi=92$
 MeV, and decay widths, $\Gamma_{\rho^0\to\mathrm{e}^+\mathrm{e}^-}=6.77$ keV,
$\Gamma_{\rho\to\pi\pi}=151$ MeV as basic requirements from low-energy QCD phenomenology.

We thus use the same parameter set as in Refs.~\cite{Blaschke:2007ce,Horvatic:2007wu},
namely $m_u=m_d=m_l=5.49$~MeV, $m_s=115$~MeV, $D_0\Lambda_0^2 = 219$, $D_1\Lambda_0^4 = 40$,
$\Lambda_0=0.758$~{GeV}, $\Lambda_1=0.961$~GeV and $p_0=0.6$~{GeV} for the rank-2 model.

For fixing the parameters in the rank-1 model, we use only the masses of pion and kaon,
 the pion decay constant, and GMOR as one additional constraint. This gives
 $m_u=m_d=m_l=6.6$~MeV, $m_s=142$~MeV, $\, D_0\Lambda_0^2 = 113.67$, and
 $\Lambda_0=0.647$~{GeV} for our values of the rank-1 parameters.


\vskip 3mm

At $T>0$, $p \to p_n = (\omega_n, {\vec p})$. Presently, pertinent are 
the fermion Matsubara frequencies $\omega_n = (2n+1)\pi T$.
Due to loss of $O(4)$ symmetry in medium, the dressed quark propagator
 (\ref{dressingS}) is at $T>0$ replaced by 
\begin{equation}
S_q^{-1}({\vec p},\omega_n;T)\, = \, i\vec{\gamma} \cdot \vec{p}\; A_q({\vec p}^2,\omega_n;T)
\, + \, i \gamma_4 \omega_n\; C_q({\vec p}^2,\omega_n;T)\, +\, B_q({\vec p}^2,\omega_n;T).\;
\label{SdressedT}
\end{equation}
For separable interactions, the dressing functions $A_q, C_q$ and $B_q$ 
depend only on the sum $p_n^2=\omega_n^2 + \vec{p}^{\,2}$.
In the separable models (\ref{sepAnsatz}), with their characteristic form
(\ref{sepSolutions}) of the propagator solutions at $T=0$, 
the dressing functions obtained as solutions of the gap equation at $T > 0$ are:
\begin{equation}
A_q(p_n^2;T) = 1 + a_q(T) {\cal F}_1(p_n^2),\quad
C_q(p_n^2;T) = 1 + c_q(T) {\cal F}_1(p_n^2), \quad
B_q(p_n^2;T) = m_q + b_q(T) {\cal F}_0(p_n^2).
\end{equation}
That is, the former gap constants $a_q$ and $b_q$ become temperature-dependent gap functions
 $a_f(T)$, $b_f(T)$ and $c_f(T)$ obtained from the nonlinear system of equations:
\begin{eqnarray}
a_q(T) = \frac{8 D_1}{9}\, T \sum_n \int \frac{d^3p}{(2\pi)^3}\,{\cal F}_1(p_n^2)\, \vec{p}^{\,2}\,
A_q(p_n^2,T)\; d_q^{-1}(p_n^2,T) \; ,
\label{oldgap1}\\
c_q(T) = \frac{8 D_1}{3}\, T \sum_n \int \frac{d^3p}{(2\pi)^3}\,{\cal F}_1(p_n^2)\, \omega_n^2\,
C_q(p_n^2,T)\; d_q^{-1}(p_n^2,T) \; ,
\label{oldgap2}\\
b_q(T) = \frac{16 D_0}{3}\, T \sum_n \int \frac{d^3p}{(2\pi)^3}\,{\cal F}_0(p_n^2)\, B_q(p_n^2,T)\;
d_q^{-1}(p_n^2,T) \; ,
\end{eqnarray}
where the denominator function is $\; d_q(p_n^2,T) = \vec{p}^{\,2}A_q^2(p_n^2,T) +
\omega_n^2\, C_q^2(p_n^2,T) + B_q^2(p_n^2,T)$.

%

%
\reftitle{references}





\end{document}